\documentclass[12pt]{amsart}
\usepackage{graphicx}

\advance\textheight\topskip
\usepackage{amsmath,amsfonts,amsthm,amssymb,amscd}
\usepackage[mathscr]{eucal}
\allowdisplaybreaks
\usepackage{hyperref}
\usepackage{times}
\usepackage[curve,matrix,arrow]{xy}

\newcommand{\CVerma}{\bar{\mathscr{V}}}

\renewcommand{\hat}{\widehat}

\newcommand{\andp}{\relax}

\newcommand{\idem}{\boldsymbol{e}}
\newcommand{\nilp}{\boldsymbol{w}}

\newcommand{\repLy}{\pi}
\newcommand{\repSw}{\bar\pi}
\newcommand{\repA}{\pi^*}

\newcommand{\bref}[1]{\textbf{\ref{#1}}}

\newcommand{\hSL}[1]{\widehat{s\ell}(#1)}

\newcommand{\polR}{\mathcal{R}}

\newcommand{\acts}{{\rightharpoondown}}

\newcommand{\cZ}{\mathfrak{Z}}

\newcommand{\Drinalg}{\mathfrak{D}}
\newcommand{\Radalg}{\mathfrak{R}}
\newcommand{\Videal}{\mathfrak{V}}
\newcommand{\Grring}{\mathfrak{G}}

\newcommand{\im}{\mathop{\mathrm{im}}\nolimits}

\newcommand{\id}{\mathrm{id}}

\renewcommand{\geq}{\,{\geqslant}\,}
\renewcommand{\leq}{\,{\leqslant}\,}

\renewcommand{\le}{\,{\leqslant}\,}

\newcommand{\Ext}{\mathrm{Ext}_{\rule{0pt}{9.5pt}%
    \overline{\mathscr{U}}_q}^1}

\newcommand{\dK}{k} 
\newcommand{\dE}{e} 
\newcommand{\ddK}{\kappa} 
\newcommand{\dF}{\phi} 
\newcommand{\coup}[2]{\langle#1,#2\rangle} 

\newcommand{\tensor}{\otimes}

\newcommand{\q}{\mathfrak{q}}
\newcommand{\qdim}{\mathop{\mathrm{qdim}}}

\newcommand{\vectv}[1]{|#1\rangle}
\newcommand{\lc}{\rep{C}}

\newcommand{\End}{\mathrm{End}}

\newcommand{\fusion}{%
  \mathop{{\otimes}\kern-7pt\raisebox{.6pt}{%
      \mbox{\footnotesize${\bullet}$}}}}

\newcommand{\floor}[1]{\lfloor#1\rfloor}

\newcommand{\UresSL}[1]{\overline{\mathscr{U}}_{\q} s\ell(#1)}

\newcommand{\UsmallSL}[1]{\mathscr{U}_{\q}s\ell(#1)^{\mathrm{small}}}
\newcommand{\mfrac}[2]{\mbox{\small$\displaystyle\frac{#1}{#2}$}}
\newcommand{\ffrac}[2]{\mbox{\footnotesize$\displaystyle\frac{#1}{#2}$}}
\newcommand{\half}{%
  \mathchoice{\ffrac{1}{2}}{\frac{1}{2}}{\frac{1}{2}}{\frac{1}{2}}}
\newcommand{\fhalf}{\ffrac{1}{2}}

\newcommand{\qbin}[2]{\mathchoice%
  {\qbinm{#1}{#2}}{\qbinmm{#1}{#2}}%
  {\qbinmm{#1}{#2}}{\qbinmm{#1}{#2}}}
\newcommand{\qbinm}[2]{\mbox{\footnotesize$\displaystyle
    \genfrac{[}{]}{0pt}{}{#1}{#2}$}}
\newcommand{\qbinmm}[2]{\genfrac{[}{]}{0pt}{}{#1}{#2}}

\newcommand{\QBIN}[2]{{\qbin{#1}{#2}}_*}

\newcommand{\angint}[1]{\langle#1\rangle}
\newcommand{\qbinom}[2]{\mathchoice%
  {\qbinomm{#1}{#2}}{\qbinommm{#1}{#2}}%
  {\qbinommm{#1}{#2}}{\qbinommm{#1}{#2}}}
\newcommand{\qbinomm}[2]{\mbox{\footnotesize$\displaystyle
    \genfrac{\langle}{\rangle}{0pt}{}{#1}{#2}$}}
\newcommand{\qbinommm}[2]{\genfrac{\langle}{\rangle}{0pt}{}{#1}{#2}}

\newcommand{\mat}[1]{\mathsf{#1}}  

\newcommand{\eigenP}{\mat{P}}

\newcommand{\cchi}{\boldsymbol{\chi}}
\newcommand{\rrho}{\boldsymbol{\rho}}
\newcommand{\hrho}{\hat{\boldsymbol{\rho}}}
\newcommand{\vvarphi}{\hat{\pmb{\varphi}}}
\newcommand{\PPhi}{\boldsymbol{\varphi}} 
\newcommand{\vvarkappa}{\pmb{\boldsymbol{\varkappa}}}
\newcommand{\llambda}{\boldsymbol{\lambda}}

\newcommand{\radmap}{\widehat{\pmb{\boldsymbol{\phi}}}}
\newcommand{\drmap}{\boldsymbol{\chi}}
\newcommand{\amap}{\boldsymbol{\xi}}

\newcommand{\modS}{\mathscr{S}}
\newcommand{\modT}{\mathscr{T}}

\newcommand{\modL}{\mathscr{P}^{+}}
\newcommand{\modP}{\mathscr{P}^{-}}
\newcommand{\repX}{\mathscr{X}}
\newcommand{\Verma}{\mathscr{V}}

\newcommand{\modQ}{\mathscr{Q}}
\newcommand{\repR}{\mathscr{R}}

\newcommand{\repLambda}{\repX^{+}}
\newcommand{\repPi}{\repX^{-}}

\newcommand{\repF}{\rep{F}}

\newcommand{\rep}{\mathscr}  

\newcommand{\Ker}{\mathop{\mathrm{Ker}}}

\newcommand{\dd}{\partial}
\newcommand{\SLiiZ}{SL(2,\oZ)}
\newcommand{\oC}{\mathbb{C}}

\newcommand{\oN}{\mathbb{N}}

\newcommand{\oZ}{\mathbb{Z}}

\newcommand{\one}{\boldsymbol{1}}
\newcommand{\tr}{\mathrm{Tr}^{\vphantom{y}}}
\newcommand{\Tr}{\mathrm{Tr}^{\vphantom{y}}}
\newcommand{\qtr}{\mathrm{qch}}
\newcommand{\qTr}{\mathrm{qCh}}
\newcommand{\ad}{\mathrm{Ad}}

\newcommand{\Ch}{\mathfrak{Ch}}

\newcommand{\pbw}{\boldsymbol{e}}
\newcommand{\dpbw}{\boldsymbol{f}}
\newcommand{\pbwd}{\boldsymbol{m}}
\newcommand{\pbwdd}{\boldsymbol{n}}

\newcommand{\toppr}{\mathsf{b}}
\newcommand{\botpr}{\mathsf{a}}
\newcommand{\leftpr}{\mathsf{x}}
\newcommand{\rightpr}{\mathsf{y}}

\newcommand{\comodul}{{\boldsymbol{a}}}
\newcommand{\coint}{{\boldsymbol{c}}}
\newcommand{\rint}{{\boldsymbol{\mu}}}

\newcommand{\balance}{{\boldsymbol{g}}}
\newcommand{\ribbon}{{\boldsymbol{v}}}
\newcommand{\sqs}{{\boldsymbol{u}}}

\newcommand{\cointa}{N}

\newcommand{\cas}{\boldsymbol{C}}

\newcommand{\cheb}{{U}}

\numberwithin{equation}{section}
\makeatletter
\@addtoreset{equation}{section}
\@addtoreset{subsubsection}{section}

\def\@secnumfont{\bfseries}
\def\subsubsection{\@startsection{subsubsection}{3}%
  \z@{.5\linespacing\@plus.7\linespacing}{-.5em}%
  {\normalfont\bfseries}}
\def\paragraph{\@startsection{paragraph}{4}%
  \z@\z@{-\fontdimen2\font}%
  \normalfont\bfseries}
\def\subparagraph{\@startsection{subparagraph}{5}%
  \z@\z@{-\fontdimen2\font}%
  \normalfont\bfseries}

\makeatother

\newcommand{\algW}{\mathcal{W}}

\swapnumbers
\newtheorem{Thm}[subsection]{Theorem}
\newtheorem{thm}[subsubsection]{Theorem}

\newtheorem{lemma}[subsubsection]{Lemma}

\newtheorem{prop}[subsubsection]{Proposition}
\newtheorem{Cor}[subsection]{Corollary}
\newtheorem{cor}[subsubsection]{Corollary}
\newtheorem{Conj}[subsection]{Conjecture}

\theoremstyle{definition}

\begin{document}

\title[Logarithmic CFTs and quantum groups]{%
  \vspace*{-4\baselineskip}
  \mbox{}\hfill\texttt{\small\lowercase{hep-th}/\lowercase{0504093}}
  \\[\baselineskip]
  Modular group representations and fusion in logarithmic
  \hbox{conformal field theories and in the quantum group center}}

\author[Feigin]{B.L.~Feigin}%

\address{\mbox{}\kern-\parindent blf: Landau Institute for Theoretical
  Physics \hfill\mbox{}\linebreak \texttt{feigin@mccme.ru}}

\author[Gainutdinov]{A.M.~Gainutdinov}%

\address{\mbox{}\kern-\parindent amg: Physics Department, Moscow State
  University \hfill\mbox{}\linebreak \texttt{azot@mccme.ru}}

\author[Semikhatov]{A.M.~Semikhatov}%

\address{\mbox{}\kern-\parindent ams, iyt: Lebedev Physics Institute
  \hfill\mbox{}\linebreak \texttt{ams@sci.lebedev.ru},
  \texttt{tipunin@td.lpi.ru}}

\author[Tipunin]{I.Yu.~Tipunin}

\begin{abstract}
  The $\SLiiZ$-representation $\repLy$ on the center of the restricted
  quantum group $\UresSL2$ at the primitive $2p$th root of unity is
  shown to be equivalent to the $\SLiiZ$-representation on the
  \textit{extended} characters of the logarithmic $(1,p)$ conformal
  field theory model.  The multiplicative Jordan decomposition of the
  $\UresSL2$ ribbon element determines the decomposition of $\repLy$
  into a ``pointwise'' product of two commuting
  $\SLiiZ$-representations, one of which restricts to the Grothendieck
  ring; this restriction is equivalent to the $\SLiiZ$-representation
  on the $(1,p)$-characters, related to the fusion algebra via a
  nonsemisimple Verlinde formula.  The Grothendieck ring of $\UresSL2$
  at the primitive $2p$th root of unity is shown to coincide with the
  fusion algebra of the $(1,p)$ logarithmic conformal field theory
  model.  As a by-product, we derive $q$-binomial identities implied
  by the fusion algebra realized in the center of~$\UresSL2$.
\end{abstract}

\maketitle


\thispagestyle{empty}


\setcounter{tocdepth}{2}

\vspace*{-24pt}

\begin{footnotesize}\addtolength{\baselineskip}{-6pt}
  \tableofcontents
\end{footnotesize}

\section{Introduction}
We study a Kazhdan--Lusztig-like correspondence between a
vertex-operator algebra and a quantum group in the case where the
conformal field theory associated with the vertex-operator algebra is
logarithmic.  In its full extent, the Kazhdan--Lusztig correspondence
comprises the following claims:
\begin{enumerate}
\item\label{item:equiv-cat} A suitable representation category of the
  vertex-operator algebra is equivalent to the category of
  finite-dimensional quantum group representations.
  
\item\label{item:Grring} The fusion algebra associated with the
  conformal field theory coincides with the quantum-group
  Grothendieck ring.
  
\item\label{item:SLiiZ} The modular group representation associated
  with conformal blocks on a torus is equivalent to the modular group
  representation on the center of the quantum group.
\end{enumerate}
Such full-fledged claims of the Kazhdan--Lusztig
correspondence~\cite{[KL]} have been established for affine Lie
algebras at a negative integer level and for some other algebras ``in
the negative zone.''  But in the positive zone, the correspondence
holds for rational conformal field models~\cite{[MS]} (such as
$(p',p)$-minimal Virasoro models and $\hSL2_k$ models with
$k\,{\in}\,\oZ_+$) with certain ``corrections.''  Notably, the
semisimple fusion in rational models corresponds to a semisimple
quasitensor category obtained as the quotient of the representation
category of a quantum group by the tensor ideal of indecomposable
tilting modules.  Taking the quotient (``neglecting the negligible''
in~\cite{[Fink]}, cf.~\cite{[T]}) makes the correspondence somewhat
indirect; in principle, a given semisimple category can thus
correspond to different quantum groups.  Remarkably, the situation is
greatly improved for the class of logarithmic (nonsemisimple) models
considered in this paper, where the quantum group itself (not only a
quasitensor category) can be reconstructed from the conformal field
theory~data.

In this paper, we are mostly interested in Claims~\ref{item:SLiiZ}
and~\ref{item:Grring}.  Claim~\ref{item:SLiiZ} of the Kazhdan--Lusztig
correspondence involves the statement that the counterpart of the
quantum group center on the vertex-operator algebra side is given by
the endomorphisms of the identity functor in the category of
vertex-operator algebra representations.  This object\,---\,morally,
the ``center'' of the associated conformal field theory\,---\,can be
identified with the finite-dimensional space $\cZ_{\mathrm{cft}}$ of
conformal blocks on a torus.  In the semisimple case,
$\cZ_{\mathrm{cft}}$ coincides with the space of conformal field
theory characters, but in the nonsemisimple case, it is not exhausted
by the characters, although we conveniently call it the (space of)
extended characters (all these are functions on the upper complex
half-plane).  The space $\cZ_{\mathrm{cft}}$ carries a modular group
representation, and the Kazhdan--Lusztig correspondence suggests
looking for its relation to the modular group representation on the
quantum group center.  We recall that an $\SLiiZ$-representation can
be defined for a class of quantum groups (in fact, for ribbon
quasitriangular categories)~\cite{[Lyu],[LM]}.  Remarkably, the two
$\SLiiZ$-representations (on $\cZ_{\mathrm{cft}}$ and on the quantum
group center~$\cZ$) are indeed equivalent for the logarithmic
conformal field theory models studied here.

The details of our study and the main results are as follows.  On the
vertex-operator algebra side, we consider the ``triplet'' W-algebra
$\algW(p)$ that was studied in~\cite{[K-first],[GK2]} in relation to
the logarithmic $(1,p)$ models of conformal field theory with
$p=2,3,\dots$.  The algebra $\algW(p)$ has $2p$ irreducible
highest-weight representations $\repX^{\pm}(s)$, $s\,{=}\,1,\dots,p$,
which (in contrast to the case of rational conformal field models)
admit nontrivial extensions among themselves ($L_0$ is
nondiagonalizable on some of extensions, which makes the theory
logarithmic).  The space $\cZ_{\mathrm{cft}}$ in the $(1,p)$-model is
$(3p\,{-}\,1)$-dimensional (cf.~\cite{[F],[F-fusion]}).

On the quantum-group side, we consider the \textit{restricted}
(``baby'' in a different nomenclature) quantum group $\UresSL2$ at the
primitive $2p$th root of unity~$\q$.  We define it in~\bref{sec:Cas}
below, and here only note the key relations
$E^{p}\,{=}\,F^{p}\,{=}\,0$, $K^{2p}\,{=}\one$ (with $K^p$ then being
central).  It has $2p$ irreducible representations and a
$(3p\,{-}\,1)$-dimensional center (Prop.~\bref{prop-center} below).
The center~$\cZ$ of~$\UresSL2$ is endowed with an
$\SLiiZ$-representation constructed as in~\cite{[Lyu],[LM],[Kerler]},
even though $\UresSL2$ is not quasitriangular~\cite{[ChP]} (the last
fact may partly explain why $\UresSL2$ is not as popular as the
\textit{small} quantum group).

\begin{Thm}\label{thm:1.1}
  The $\SLiiZ$-representations on $\cZ_{\mathrm{cft}}$ and on~$\cZ$
  are equivalent.
\end{Thm}
Thus, Claim~\ref{item:SLiiZ} of the Kazhdan--Lusztig correspondence is
fully valid for $\algW(p)$ and $\UresSL2$
at~$\q\,{=}\,e^{\frac{i\pi}{p}}$.  We let $\repLy$ denote the
$\SLiiZ$-representation in the theorem.

\medskip

Regarding Claim~\ref{item:Grring}, we first note that, strictly
speaking, the fusion for $\algW(p)$, understood in its ``primary''
sense of calculation of the coinvariants, has been derived only for
$p\,{=}\,2$~\cite{[GK1]}.  In rational conformal field theories, the
Verlinde formula~\cite{[V]} allows recovering fusion from the modular
group action on characters.  In the $(1,p)$ logarithmic models, the
procedure proposed in~\cite{[FHST]} as a nonsemisimple generalization
of the Verlinde formula allows constructing a commutative associative
algebra from the $\SLiiZ$-action on the
  $\algW(p)$-characters.  This algebra $\Grring_{2p}$ on $2p$
elements $\chi^{\alpha}(s)$ ($\alpha\,{=}\,\pm1$, $s\,{=}\,1,\dots,p$)
is given~by
\begin{equation}\label{the-fusion}
  \chi^{\alpha}(s)\chi^{\alpha'}(s')
  =\sum_{\substack{s''=|s - s'| + 1\\
      \mathrm{step}=2}}^{s + s' - 1}
  \tilde\chi^{\alpha\alpha'}(s'')
\end{equation}
where
\begin{equation*}
  \tilde\chi^{\alpha}(s)
  =
  \begin{cases}
    \chi^{\alpha}(s),&1\leq s\leq p,\\
    \chi^{\alpha}( 2p - s) + 2\chi^{-\alpha}( s - p), & p + 1 \leq s
    \leq 2p - 1.
  \end{cases}
\end{equation*}
For $p\,{=}\,2$, this algebra coincides with the fusion
in~\cite{[GK1]}, and we believe that it is indeed the fusion for
all~$p$.  Our next result in this paper strongly supports this claim,
setting it in the framework of the Kazhdan--Lusztig correspondence
between $\algW(p)$ and~$\UresSL2$ at~$\q=e^{\frac{i\pi}{p}}$.

\begin{Thm}\label{thm:1.2}
  Let $\q=e^{\frac{i\pi}{p}}$.  Under the identification of
  $\chi^{\alpha}(s)$, $\alpha=\pm1$, $s\,{=}\,1,\dots, p$, with
  the $2p$ irreducible $\UresSL2$-representations, the algebra
  $\Grring_{2p}$ in~\eqref{the-fusion} is the Grothen\-dieck ring
  of~$\;\UresSL2$.

\end{Thm}
We emphasize that the algebras are isomorphic as fusion algebras,
i.e., including the identification of the respective preferred bases
given by the irreducible representations.

The procedure in~\cite{[FHST]} leading to fusion~\eqref{the-fusion} is
based on the following structure of the
$\SLiiZ$-representation~$\repLy$ on~$\cZ_{\mathrm{cft}}$ in the
$(1,p)$ model:
\begin{equation}\label{repLy-struct}
  \cZ_{\mathrm{cft}}
  =\mathcal{R}_{p+1}\oplus\oC^2\tensor\mathcal{R}_{p-1}.
\end{equation}
Here, $\mathcal{R}_{p+1}$ is a $(p\,{+}\,1)$-dimensional
$\SLiiZ$-representation (actually, on characters of a lattice
vertex-operator algebra), \ $\mathcal{R}_{p-1}$ is a
$(p\,{-}\,1)$-dimen\-sional $\SLiiZ$-representation (actually, the
representation on the unitary $\hSL2_{k}$-characters at the level
$k\,{=}\,p\,{-}\,2$), and $\oC^2$ is the standard two-dimensional
$\SLiiZ$-representation.  Equivalently,~\eqref{repLy-struct} is
reformulated as follows.  We have two $\SLiiZ$-representations
$\repSw$ and~$\repA$ on~$\cZ_{\mathrm{cft}}$ in terms of which
$\repLy$ factors as $\repLy(\gamma)\,{=}\,\repA(\gamma)\repSw(\gamma)$
$\forall\gamma\in\SLiiZ$ and which commute with each other,
$\repA(\gamma)\repSw(\gamma')=\repSw(\gamma')\repA(\gamma)$; moreover,
$\repSw$ restricts to the $2p$-dimensional space of the
$\algW(p)$-characters.

In view of Theorem~\bref{thm:1.1}, this structure of the
$\SLiiZ$-representation is reproduced on the quantum-group side: there
exist $\SLiiZ$-representations $\repSw$ and $\repA$\pagebreak[3] on the
center~$\cZ$ of $\UresSL2$ in terms of which the representation
in~\cite{[Lyu],[LM]} factors.  Remarkably, these
representations~$\repSw$ and~$\repA$ on~$\cZ$ can be constructed in
intrinsic quantum-group terms, by modifying the construction
in~\cite{[Lyu],[LM]}.  We recall that the $\modT$ generator of
$\SLiiZ$ is essentially given by the ribbon element~$\ribbon$, and the
$\modS$ generator is constructed as the composition of the Radford and
Drinfeld mappings.  That~$\repSw$ and~$\repA$ exist is related to the
multiplicative Jordan decomposition of the ribbon element
$\ribbon\,{=}\,\bar\ribbon\ribbon^*$, where~$\bar\ribbon$ is the
semisimple part and~$\ribbon^*$ is the unipotent (one-plus-nilpotent)
part.  Then $\bar\ribbon$ and $\ribbon^*$ yield the respective
``$T$''-generators $\bar\modT$ and $\modT^*$.  The corresponding
``$S$''-generators $\bar\modS$ and~$\modS^*$ are constructed by
deforming the Radford and Drinfeld mappings \textit{respectively}, as
we describe in Sec.~\bref{two-rep-on-Z} below.  We temporarily call
the $\SLiiZ$-representations $\repSw$ and~$\repA$ the representations
\textit{associated with} $\bar\ribbon$ and $\ribbon^*$.

\begin{Thm}\label{thm:modular-2}
  Let $\ribbon\,{=}\,\bar\ribbon\ribbon^*$ be the multiplicative
  Jordan decomposition of the $\UresSL2$ ribbon element (with
  $\bar\ribbon$ being the semisimple part) and let $\repSw$
  and~$\repA$ be the respective $\SLiiZ$-representations on $\cZ$
  associated with $\bar\ribbon$ and $\ribbon^*$.  Then
  \begin{enumerate}
  \item $\repSw(\gamma)\repA(\gamma')
    \,{=}\,\repA(\gamma')\repSw(\gamma)$ for all
    $\gamma,\gamma'\in\SLiiZ$,
    
  \item $\repLy(\gamma)\,{=}\,\repSw(\gamma)\repA(\gamma)$ for all
    $\gamma\in\SLiiZ$, and
    
  \item the representation $\repSw$ restricts to the image of the
    Grothendieck ring in the center.
  \end{enumerate}
\end{Thm}
The image of the Grothendieck ring in this theorem is under the
Drinfeld mapping.  The construction showing how the
representations~$\repSw$ and~$\repA$ on the center are derived from
the Jordan decomposition of the ribbon element is developed in
Sec.~\bref{two-rep-on-Z} only for $\UresSL2$, but we expect it to be
valid in general.
\begin{Conj}
  The multiplicative Jordan decomposition of the ribbon element gives
  rise to $\SLiiZ$-representations $\repSw$ and~$\repA$ with the
  properties as in Theorem~\bref{thm:modular-2} for any factorizable
  ribbon quantum group.
\end{Conj}

Regarding Claim~\ref{item:equiv-cat} of the Kazhdan--Lusztig
correspondence associated with the $(1,p)$ logarithmic models, we only
formulate a conjecture; we expect to address this issue in the future,
beginning with~\cite{[FGST2]}, where, in particular, the
representation category is studied in great detail.  In a sense, the
expected result is more natural than in the semisimple$/$rational case
because (as in Theorem~\bref{thm:1.2}) it requires no
``semisimplification'' on the quantum-group side.
\begin{Conj}
  The category of $\algW(p)$-representations is equivalent to the
  category of finite-dimensional $\UresSL2$-representations
  with~$\q\,{=}\,e^{\frac{i\pi}{p}}$.
\end{Conj}

{}From the reformulation of fusion~\eqref{the-fusion} in quantum-group
terms (explicit evaluation of the product in the image of the
Grothendieck ring in the center under the Drinfeld mapping), we obtain
a combinatorial corollary of Theorem~\bref{thm:1.2} (see~\eqref{q-bin}
for the notation regarding $q$-binomial coefficients):
\begin{Cor}\label{lemma:the-identity}
  For $s+s'\geq n\geq m\geq0$, there is the $q$-binomial identity
  \begin{multline}\label{the-identity}
    \sum_{j\in\oZ}\sum_{i\in\oZ} q^{2 m i + j (2 n - s - s') + m s}
    \qbin{n - i}{j} \qbin{i}{m - j} \qbin{i + j + s - n}{j}
    \qbin{m - i - j + s'}{m - j}={}\\
    {}= q^{2 m n} \sum_{\ell=0}^{\min(s, s')} \qbin{n - \ell}{m}
    \qbin{m + s + s' - \ell - n}{m}.
  \end{multline}
\end{Cor}
The multiplication in algebra~\eqref{the-fusion}, which underlies this
identity, is alternatively characterized in terms of Chebyshev
polynomials, see~\bref{prop:quotient} below.
  
\medskip

There are numerous relations to the previous work.  The fundamental
results in~\cite{[Lyu],[LM]} regarding the modular group action on the
center of a Drinfeld double can be ``pushed forward'' to~$\UresSL2$,
which is a ribbon quantum group.  We note that in the standard
setting~\cite{[RT]}, a ribbon Hopf algebra is assumed to be
quasitriangular.  This is not the case with $\UresSL2$, but we keep
the term ``ribbon'' with the understanding that $\UresSL2$ is a
subalgebra in a quasitriangular Hopf algebra from which it inherits
the ribbon structure, as is detailed in what follows.  The
structure~\eqref{repLy-struct}, already implicit in~\cite{[FHST]}, is
parallel to the property conjectured in~\cite{[Kerler]} for the
$\SLiiZ$-representation on the center of the \textit{small} quantum
group $\UsmallSL2$.  Albeit for a different quantum group, we extend
the argument in~\cite{[Kerler]} by choosing the bases in the center
that lead to a simple proof and by giving the underlying Jordan
decomposition of the ribbon element and the corresponding deformations
of the Radford and Drinfeld mappings.  The $(3p\,{-}\,1)$-dimensional
center of~$\UresSL2$ at $\q$ the primitive $2p$th root of unity is
twice as big as the center of~$\UsmallSL2$ for~$\q$ the primitive
$p$th root of unity (for odd~$p$)~\cite{[Kerler],[L-center]}.  We
actually find the center of~$\UresSL2$ by studying the bimodule
decomposition of the regular representation (the decomposition of
$\UsmallSL2$ under the \textit{adjoint} action has been the subject of
some interest; see~\cite{[Ostrik]} and the references therein).  There
naturally occur indecomposable $2p$-dimensional
$\UresSL2$-representations (projective modules), which have also
appeared in~\cite{[RT],[GL],[JMT]}.  On the conformal field theory
side, the $\algW(p)$ algebra was originally studied
in~\cite{[K-first],[GK2]}, also
see~\cite{[G-alg],[F-bits]}.\enlargethispage{\baselineskip}

This paper can be considered a continuation (or a quantum-group
counterpart) of~\cite{[FHST]} and is partly motivated by remarks
already made there.  That the quantum dimensions of the irreducible
$\algW(p)$-representations are dimensions of quantum-group
representations was noted in~\cite{[FHST]} as an indication of a
quantum group underlying the fusion algebra derived there.  For the
convenience of the reader, we give most of the necessary reference
to~\cite{[FHST]} in Sec.~\ref{sec:CFTetc} and recall the crucial
conformal field theory formulas there.\footnote{We note a minor
  terminological discrepancy: in~\cite{[FHST]}, the ``fusion'' basis
  (the one with nonnegative integer structure coefficients) was called
  canonical, while in this paper we call it the preferred basis,
  reserving ``canonical'' for the basis of primitive idempotents and
  elements in the radical.}  \ In Sec.~\ref{sec:all-Usl2}, we define
the restricted quantum group $\UresSL2$, describe some classes of its
representations (most importantly, irreducible), and find its
Grothendieck ring.  In Sec.~\ref{sec:new}, we collect the
facts pertaining to the ribbon structure and the structure of a
factorizable Hopf algebra on~$\UresSL2$.  There, we also find the center
of $\UresSL2$ in rather explicit terms.  In
Sec.~\ref{sec:SLiiZ-restr}, we study $\SLiiZ$-representations on the
center of $\UresSL2$ and establish the equivalence to the
representation in~Sec.~\ref{sec:CFTetc} and the factorization
associated with the Jordan decomposition of the ribbon element.

The Appendices contain auxiliary or bulky material.  In
Appendix~\ref{app:Hopf}, we collect a number of standard facts about
Hopf algebras that we use in the paper.  In Appendix~\ref{sec:double},
we construct a Drinfeld double that we use to derive the $M$-matrix
and the ribbon element for~$\UresSL2$.  In
Appendix~\ref{verma-proj-mod-base}, we give the necessary details
about indecomposable $\UresSL2$-modules.  The ``canonical'' basis in
the center of $\UresSL2$ is explicitly constructed in
Appendix~\ref{app:center}.  As an elegant corollary of the description
of the Grothendieck ring in terms of Chebyshev polynomials, we
reproduce the formulas for the eigenmatrix in~\cite{[FHST]}.
Appendix~\ref{app:derivation} is just a calculation leading to
identity~\eqref{the-identity}.

\subsection*{Notation} We use the standard notation
\begin{equation*}
  [n] = \ffrac{q^n-q^{-n}}{q-q^{-1}},\quad
  n\in\oZ,\quad [n]! = [1][2]\dots[n],\quad n\in\oN,\quad[0]!=1
\end{equation*}
(without indicating the ``base''~$q$ explicitly) and set
\begin{equation}\label{q-bin}
  \qbin{m}{n}=
  \begin{cases}
    0,& n<0\quad\text{or}\quad m-n<0,\\
    \ffrac{[m]!}{[n]!\,[m-n]!}&\text{otherwise}.
  \end{cases}
\end{equation}

In referring to the root-of-unity case, we set
\begin{equation*}
  \q=e^{\frac{i\pi}{p}}
\end{equation*}
for an integer $p\geq2$.  The $p$ parameter is as in
Sec.~\bref{sec:CFTetc}.

For Hopf algebras in general (in the Appendices) and for $\UresSL2$
specifically, we write $\Delta$, $\epsilon$, and~$S$ for the
comultiplication, counit, and antipode respectively.  Some other
conventions are as follows:
\begin{itemize}
  
\item[$\cZ$] --- the quantum group center,
  
\item[$\Ch$] --- the space of $q$-characters (see~\bref{sec:q-chars}),
  
\item[$\rint$] --- the integral (see~\bref{app:int}),
  
\item[$\coint$] --- the cointegral (see~\bref{app:int}),
  
\item[$\balance$] --- the balancing element (see~\bref{app:int}),
  
\item[$\ribbon$] --- the ribbon element (see~\bref{sec:ribbon}),
  
\item[$\bar M$] --- the $M$-matrix (see~\bref{app:M}; $\bar M$ is used
  for $\UresSL2$ and $M$ in general),

\item[$\drmap$] --- the Drinfeld mapping $A^*\to A$
  (see~\bref{sec:Drpdef}),
  
\item[$\cchi^{\pm}(s)$] --- the image of the irreducible
  $\UresSL2$-representation $\repX^{\pm}(s)$ in the center under the
  Drinfeld mapping (see~\bref{fusion-center}),
  
\item[$\radmap$] --- the Radford mapping $A^*\to A$
  (see~\bref{sec:radford-all}),
  
\item[$\radmap{}^{\pm}(s)$] --- the image of the irreducible
  $\UresSL2$-representation $\repX^{\pm}(s)$ in the center under the
  Radford mapping (see~\bref{sec:Radford-SL2}),
  
\item[$\repX^{\pm}(s)$] --- irreducible $\UresSL2$-representations
  (see~\bref{subsec:irrep}); in~\bref{sec:VOA}, irreducible
  $\algW(p)$-rerpe\-sentations.

\item[$\Verma^{\pm}(s)$] --- Verma modules
  (see~\bref{subsec:verma-mod} and~\bref{verma-mod-base}),
  
\item[$\CVerma^{\pm}(s)$] --- contragredient Verma modules
  (see~\bref{verma-mod-base}),
  
\item[$\mathscr{P}^{\pm}(s)$] --- projective $\UresSL2$-modules
  (see~\bref{subsec:proj-mod} and~\bref{proj-mod-base}),
  
\item[$\qTr_{\repX}$] --- the $q$-character of a
  $\UresSL2$-representation~$\repX$ (see~\bref{app:qCh}),
  
\item[$\Grring_{2p}$] --- the $\UresSL2$ Grothendieck ring;
  $\Grring(A)$ is the Grothendieck ring of a Hopf algebra~$A$,
  
\item[$\Drinalg_{2p}$] --- the Grothendieck ring image in the center
  under the Drinfeld mapping,
  
\item[$\Radalg_{2p}$] --- the Grothendieck ring image in the center
  under the Radford mapping.
\end{itemize}

We write $x'$, $x''$, $x'''$, etc.\ (Sweedler's notation) in
constructions like
\begin{equation*}
  \Delta(x)=\sum_{(x)}x'\tensor x'',\quad
  (\Delta\tensor\id)\Delta(x)=\sum_{(x)}x'\tensor x''\tensor x''',
  \quad\dots.
\end{equation*}
For a linear function $\beta$, we use the notation $\beta(?)$, where
$?$ indicates the position of its argument in more complicated
constructions.

We choose two elements generating $\SLiiZ$ as
$\left(\begin{smallmatrix}
    0&1\\
    -1&0
  \end{smallmatrix}\right)$  and $\left(\begin{smallmatrix}
    1&1\\
    0&1
  \end{smallmatrix}\right)$ and use  the notation of the
type $\modS$, $\modS^*$, $\bar\modS$, \dots\ and $\modT$, $\modT^*$,
$\bar\modT$, \dots\ for these elements in various representations.

\section{Vertex-operator algebra for the $(1,p)$-conformal field
  theory,\\* its characters, and $\SLiiZ$-representations}
\label{sec:CFTetc} Logarithmic models of conformal field theory, of
which the $(1,p)$-models are an example, were introduced
in~\cite{Gurarie} and were considered, in particular,
in~\cite{[GK1],[GK2],Roh,[F],[G-alg],[F-bits],[FFHST],[FHST],[GurLu]}
(also see the references therein).  Such models are typically defined
as kernels of certain screening operators.  The actual symmetry of the
theory is the maximal local algebra in this kernel.  In the
$(1,p)$-model, which is the kernel of the ``short'' screening
operator, see~\cite{[FHST]}, this is the W-algebra $\algW(p)$ studied
in~\cite{[K-first],[GK2]}.  We briefly recall it in~\bref{sec:VOA}.
In~\bref{mod-on-char}, we give the modular transformation properties
of the $\algW(p)$-characters and identify the
$(3p\,{-}\,1)$-dimensional $\SLiiZ$-representation
on~$\cZ_{\mathrm{cft}}$ (the space of extended characters).
In~\bref{thm:R-decomp}, we describe the structure of this
representation.

\subsection{VOA}\label{sec:VOA}
Following~\cite{[FHST]}, we consider the vertex-operator algebra
$\algW(p)$\,---\,the W-al\-gebra studied in~\cite{[K-first],[GK2]},
which can be described in terms of a single free field $\varphi(z)$
with the operator product expansion $\varphi(z)\,\varphi(w)=
\log(z-w)$.  For this, we introduce the energy-momentum tensor
\begin{equation}\label{eq:the-Virasoro}
  T=\fhalf\,\dd\varphi\,\dd\varphi+\ffrac{\alpha_0}{2}\,\dd^2\varphi,
  \quad
  \alpha_+=\sqrt{2p},\quad\alpha_-=-\sqrt{\ffrac{2}{p}},
  \quad
  \alpha_0=\alpha_++\alpha_-,
\end{equation}
with central charge $c\,{=}\,13\,{-}\,6(p\,{+}\,\frac{1}{p})$, and the
set of vertex operators $V_{r,s}(z)=e^{j(r,s)\varphi}(z)$ with
$j(r,s)=\frac{1-r}{2}\alpha_++\frac{1-s}{2}\alpha_-$.  Let $\repF$ be
the sum of Fock spaces corresponding to $V_{r,s}(z)$ for~ $r\in\oZ$
and $1\leq s\leq p$ (see the details in~\cite{[FHST]}).  There exist
two screening operators
\begin{equation*}
  S_+=\oint e^{\alpha_+\varphi},\qquad
  S_-=\oint e^{\alpha_-\varphi},
\end{equation*}
satisfying $[S_{\pm},T(z)]\,{=}\,0$.  We define $\algW(p)$ as a
maximal local subalgebra in the kernel of the ``short''
screening~$S_-$.  The algebra $\algW(p)$ is generated by the currents
\begin{equation*}
  W^-(z)=e^{-\alpha_+\varphi}(z),\quad\;
  W^0(z)=[S_+,W^-(z)],\quad\;
  W^+(z)=[S_+,W^0(z)]
\end{equation*}
(which are primary fields of dimension~$2p\,{-}\,1$ with respect to
energy-momentum tensor~\eqref{eq:the-Virasoro}).  The algebra
$\algW(p)$ has $2p$ irreducible highest-weight representations,
denoted as $\repX^{+}(s)$ and $\repX^{-}(s)$, $1\leq s\leq p$ (the
respective representations $\mathit{\Lambda}(s)$ and
$\mathit{\Pi}\!(s)$ in~\cite{[FHST]}).  The highest-weight vectors in
$\repX^{+}(s)$ and $\repX^{-}(s)$ can be chosen as $V_{0,s}$ and
$V_{1,s}$ respectively.

It turns out that
\begin{equation*}
  \Ker S_-\!\!\Bigm|_{\repF}=
  \bigoplus_{s=1}^{p}\repLambda(s)\oplus\repPi(s).
\end{equation*}

\subsection{$\algW(p)$-algebra characters and the
  $\SLiiZ$-representation on~$\cZ_{\mathrm{cft}}$}\label{mod-on-char}
We now recall~\cite{[FHST]} the modular transformation properties of
the $\algW(p)$-characters
\begin{equation*}
  \chi_{s\andp}^{+}(\tau)
  =\Tr_{\repX^+(s)}e^{2i\pi\tau(L_0-\frac{c}{24})}, 
  \qquad\chi_{s\andp}^{-}(\tau)
  =\Tr_{\repX^-(s)}e^{2i\pi\tau(L_0-\frac{c}{24})},
  \qquad
  1\leq s\leq p
\end{equation*}
(the respective characters~$\chi_{s,p}^{\mathit{\Lambda}}(\tau)$
and~$\chi_{s,p}^{\mathit{\Pi}}(\tau)$ in~\cite{[FHST]}), where $L_0$
is a Virasoro generator, the zero mode of energy-momentum
tensor~\eqref{eq:the-Virasoro}.  Under the $\modS$-transformation
of~$\tau$, these characters transform as
\begin{multline}\label{S-chi-plus}
  \chi_{s\andp}^{+}(-\ffrac{1}{\tau}) =\ffrac{1}{\sqrt{2p}}\biggl(
  \ffrac{s}{p}\,\Bigl[ \chi_{p\andp}^{+}(\tau) +
  (-1)^{p-s}\chi_{p\andp}^{-}(\tau)\\*
  + \sum_{s'=1}^{p-1}
  \q^{(p-s)s'}_{+} \bigl(\chi_{p-s'\andp}^{+}(\tau)
  + \chi_{s'\andp}^{-}(\tau)\bigr) \Bigr] -
  \sum_{s'=1}^{p-1}(-1)^{p+s+s'}\q^{ss'}_{-} \varphi_{s'\andp}(\tau)
  \biggr)
\end{multline}
and
\begin{multline}\label{S-chi-minus}
  \chi_{s\andp}^{-}(-\ffrac{1}{\tau}) =\ffrac{1}{\sqrt{2p}}\biggl(
  \ffrac{s}{p}\Bigl[
  \chi_{p\andp}^{+}(\tau) + (-1)^{s}\chi_{p\andp}^{-}(\tau)\\*
  + \sum_{s'=1}^{p-1}\q^{s s'}_{+} \bigl(\chi_{p-s'\andp}^{+}(\tau) +
  \chi_{s'\andp}^{-}(\tau)\bigr) \Bigr] +
  \smash[b]{\sum_{s'=1}^{p-1}}(-1)^{s+1} \q^{s' s}_{-}
  \varphi_{s'\andp}(\tau)\biggr),
\end{multline}
where $\q^{ss'}_{\pm} = \q^{ss'} \pm \q^{-ss'}$, $\q=e^{i\pi/p}$, and
we introduce the notation
\begin{equation}\label{varphi}
  \varphi_{s\andp}(\tau)=
  \tau\bigl(\ffrac{p-s}{p}\,\chi_{s\andp}^{+}(\tau) -
  \ffrac{s}{p}\,\chi_{p-s\andp}^{-}(\tau)\bigr),
  \quad 1\leq s\leq p-1.
\end{equation}

The $\algW(p)$-characters are in fact combinations of modular forms of
different weights, and hence their modular transformations involve
explicit occurrences of~$\tau$; in the formulas above, $\tau$ enters
only linearly, but much more complicated functions of~$\tau$ (and
other arguments of the characters) can be involved in nonrational
theories, cf.~\cite{[STT]}.  In the present case, because of the
explicit occurrences of~$\tau$, the $\SLiiZ$-representation space
turns out to be $(3p\,{-}\,1)$-dimensional, spanned by
$\chi^{\pm}_s(\tau)$, $1\leq s\leq p$, and $\varphi_{s\andp}(\tau)$,
$1\leq s\leq p-1$.  Indeed, we have
\begin{equation}\label{Sw-varphi}
  \varphi_{s\andp}(-\ffrac{1}{\tau})
  =\ffrac{1}{\sqrt{2p}}\sum_{s'=1}^{p-1}(-1)^{p+s+s'}
  \q^{s s'}_{-}\rho_{s'\andp}(\tau),
\end{equation}
where for the future convenience we introduce a special notation for
certain linear combinations of the characters:
\begin{equation}\label{rho}
  \rho_{s\andp}(\tau)=
  \ffrac{p-s}{p}\,\chi_{s\andp}^{+}(\tau) -
  \ffrac{s}{p}\,\chi_{p-s\andp}^{-}(\tau),
  \quad 1\leq s\leq p-1.
\end{equation}

Under the $\modT$-transformation of~$\tau$, the $\algW(p)$-characters
transform as
\begin{equation}\label{T-chi}
  \chi_{s\andp}^{+}(\tau+1) =
  \lambda_{p,s}\chi_{s\andp}^{+}(\tau),
  \quad
  \chi_{p-s\andp}^{-}(\tau+1) =
  \lambda_{p,s}
  \chi_{p-s\andp}^{-}(\tau),
\end{equation}
where 
\begin{equation}\label{eq:lambda}
  \lambda_{p,s}=e^{i\pi(\frac{(p-s)^2}{2p}-\frac{1}{12})},
\end{equation}
and hence
\begin{equation}\label{Tsw-varphi}
  \varphi_{s\andp}(\tau+1) =
  \lambda_{p,s}
  \bigl(\varphi_{s\andp}(\tau) + \rho_{s\andp}(\tau)\bigr).
\end{equation}

We let $\cZ_{\mathrm{cft}}$ denote this $(3p\,{-}\,1)$-dimensional
space spanned by $\chi^{\pm}_s(\tau)$, $1\leq s\leq p$, and
$\varphi_{s\andp}(\tau)$, $1\leq s\leq p-1$.  As noted in the
introduction, $\cZ_{\mathrm{cft}}$ is the space of conformal blocks on
the torus, which is in turn isomorphic to the endomorphisms of the
identity functor.  Let $\repLy$ be the $\SLiiZ$-representation on
$\cZ_{\mathrm{cft}}$ defined by the above formulas.

\begin{Thm}\label{thm:R-decomp}
  \addcontentsline{toc}{subsection}{\thesubsection. \ \ Factorization
    of $\repLy$} The $\SLiiZ$-representation on $\cZ_{\mathrm{cft}}$
  has the structure
  \begin{equation*}
    \cZ_{\mathrm{cft}}
    =\mathcal{R}_{p+1}\oplus\oC^2\tensor\mathcal{R}_{p-1},
  \end{equation*}
  where $\mathcal{R}_{p+1}$ and $\mathcal{R}_{p-1}$ are
  $\SLiiZ$-representations of the respective dimensions $p\,{+}\,1$
  and $p\,{-}\,1$, and $\oC^2$ is the two-dimensional representation.
  This implies that there exist $\SLiiZ$-representations $\repSw$ and
  $\repA$ on $\cZ_{\mathrm{cft}}$ such that
  \begin{equation*}
    \repLy(\gamma)=\repA(\gamma)\repSw(\gamma),\quad
    \repSw(\gamma)\repA(\gamma') =\repA(\gamma')\repSw(\gamma),
    \qquad
    \gamma,\gamma' \in\SLiiZ.
  \end{equation*}
\end{Thm}
\begin{proof}
  Let $\mathcal{R}_{p+1}$ be spanned by
  \begin{equation}\label{varkappa}
    \begin{split}
      \varkappa_0(\tau) &= \chi_{p\andp}^{-}(\tau),\\
      \varkappa_{s\andp}(\tau) &= \chi^{+}_{s\andp}(\tau) +
      \chi^{-}_{p-s\andp}(\tau),
      \quad 1\leq s\leq p-1,\\
      \varkappa_p(\tau) &= \chi_{p\andp}^{+}(\tau)
    \end{split}
  \end{equation}
  (these are the characters of \textit{Verma} modules
  over~$\algW(p)$).  The formulas in~\bref{mod-on-char} show that
  $\mathcal{R}_{p+1}$ is an $\SLiiZ$-representation; namely, it
  follows that
  \begin{gather*}
    \modT\varkappa_{s}(\tau)
    = \lambda_{p,s}\varkappa_{s}(\tau)
  \end{gather*}
  and
  \begin{equation*}
    \modS\varkappa_{s\andp}(\tau)=\hat\varkappa_{s\andp}(\tau),
    \qquad
    \modS\hat\varkappa_{s\andp}(\tau)=\varkappa_{s\andp}(\tau),
  \end{equation*}
  where
  \begin{equation*}
    \hat\varkappa_{s\andp}(\tau)
    =\ffrac{1}{\sqrt{2p}}
    \bigl((-1)^{p-s}\varkappa_{0\andp}(\tau) +
    \sum_{s'=1}^{p-1}
    (-1)^{s'}\q^{s s'}_{+}\varkappa_{p-s'\andp}(\tau)
    + \varkappa_{p\andp}(\tau)\bigr),
    \quad 0\leq s\leq p,
  \end{equation*}
  is another basis in~$\mathcal{R}_{p+1}$.  
  
  Next, let $\mathcal{R}'_{p-1}$ be the space spanned by
  $\varphi_{s\andp}(\tau)$ in~\eqref{varphi}; another basis in
  $\mathcal{R}'_{p-1}$ is
  \begin{equation*}
    \hat\varphi_{s\andp}(\tau)=
    -\ffrac{1}{\sqrt{2p}}\sum_{s'=1}^{p-1}(-1)^{p+s+s'}
    \q^{s s'}_{-}\varphi_{s'\andp}(\tau),
    \quad 1\leq s\leq p-1.
  \end{equation*}
  Finally, let another $(p-1)$-dimensional space $\mathcal{R}''_{p-1}$
  be spanned by $\rho_{s\andp}(\tau)$ in~\eqref{rho}; another basis in
  $\mathcal{R}''_{p-1}$ is given by
  \begin{gather*}   
    \hat\rho_{s\andp}(\tau)
    =\ffrac{1}{\sqrt{2p}}\sum_{s'=1}^{p-1}(-1)^{p+s+s'} \q^{s
      s'}_{-}\rho_{s'\andp}(\tau), \quad 1\leq s\leq p-1.
  \end{gather*}  
  Equations~\eqref{S-chi-plus}--\eqref{Sw-varphi} then imply that
  \begin{alignat*}{2}    
    \modS\varphi_{s\andp}(\tau)&=\hat\rho_{s\andp}(\tau),&\qquad
    \modS\hat\varphi_{s\andp}(\tau)&=\rho_{s\andp}(\tau),\\
    \modS\rho_{s\andp}(\tau)&=\hat\varphi_{s\andp}(\tau),&
    \modS\hat\rho_{s\andp}(\tau)&=\varphi_{s\andp}(\tau),
  \end{alignat*}
  and the $\modT$-transformations in
  Eqs.~\eqref{T-chi}--\eqref{Tsw-varphi} are expressed as
  \begin{gather*}\label{T-blocks}    
    \modT
    \begin{pmatrix}
      \rho_{s}(\tau)\\
      \varphi_{s}(\tau)
    \end{pmatrix}=
    \lambda_{p,s} \,
    \begin{pmatrix}
      1 & 0\\
      1 & 1
    \end{pmatrix}
    \begin{pmatrix}
      \rho_{s}(\tau)\\
      \varphi_{s}(\tau)
    \end{pmatrix},
    \quad 1\leq s\leq p{-}1.
  \end{gather*}
  Therefore, the representation $\repLy$ has the structure
  $\mathcal{R}_{p+1}\oplus\oC^2\tensor\mathcal{R}_{p-1}$, where
  $\oC^2\tensor\mathcal{R}_{p-1}$ is spanned by
  $(\varphi_{s\andp}(\tau),\rho_{s\andp}(\tau))$, $1\leq s\leq p-1$.
  
  We now let $\bar\modS\equiv\repSw(\left(\begin{smallmatrix}
      0&1\\
      -1&0
    \end{smallmatrix}\right))$ and
  $\modS^*\equiv\repA(\left(\begin{smallmatrix}
      0&1\\
      -1&0
    \end{smallmatrix}\right))$ act on $ \cZ_{\mathrm{cft}}$ as
  \begin{alignat*}{4}
    \bar\modS\varkappa_{s\andp}(\tau)&=\hat\varkappa_{s\andp}(\tau),
    &\quad
    \bar\modS\varphi_{s\andp}(\tau)&=\hat\varphi_{s\andp}(\tau),
    &\quad
    \bar\modS\rho_{s\andp}(\tau)&=-\hat\rho_{s\andp}(\tau),\\
    \bar\modS\hat\varkappa_{s\andp}(\tau)&=\varkappa_{s\andp}(\tau),
    &\quad
    \bar\modS\hat\varphi_{s\andp}(\tau)&=-\varphi_{s\andp}(\tau),&
    \bar\modS\hat\rho_{s\andp}(\tau)&=\rho_{s\andp}(\tau),\\
    \modS^*\varkappa_{s\andp}(\tau)&=\varkappa_{s\andp}(\tau),
    \quad&
    \modS^*\varphi_{s\andp}(\tau)&=-\rho_{s\andp}(\tau),&
    \modS^*\hat\rho_{s\andp}(\tau)&=-\hat\varphi_{s\andp}(\tau),\\
    \modS^*\hat\varkappa_{s\andp}(\tau)&=\hat\varkappa_{s\andp}(\tau),&
    \modS^*\rho_{s\andp}(\tau)&=\varphi_{s\andp}(\tau),&
    \modS^*\hat\varphi_{s\andp}(\tau)&=\hat\rho_{s\andp}(\tau).
  \end{alignat*}
  and let $\bar\modT\equiv\repSw(\left(\begin{smallmatrix}
      1&1\\
      0&1
    \end{smallmatrix}\right))$ and
  $\modT^*\equiv\repA(\left(\begin{smallmatrix}
      1&1\\
      0&1
    \end{smallmatrix}\right))$ act as
  \begin{gather*}
    \bar\modT\varkappa_s(\tau)=\lambda_{p,s}
    \varkappa_s(\tau),\qquad 0\leq s\leq p,
    \\
    \bar\modT
    \begin{pmatrix}
      \rho_{s}(\tau)\\
      \varphi_{s}(\tau)
    \end{pmatrix}
    = \lambda_{p,s}
    \begin{pmatrix}
      1 \, &0\\
      0 & 1
    \end{pmatrix}
    \begin{pmatrix}
      \rho_{s}(\tau)\\
      \varphi_{s}(\tau)
    \end{pmatrix},
    \qquad 1\leq s\leq p-1,
  \end{gather*}
  and
  \begin{gather*}
    \modT^*\varkappa_s(\tau)=\varkappa_s(\tau),\qquad 0\leq s\leq p,
    \\
    \modT^*\begin{pmatrix}
      \rho_s(\tau)\\
      \varphi_s(\tau)
    \end{pmatrix}=
    \begin{pmatrix}
      1 & 0\\
      1 & 1\\
    \end{pmatrix}
    \begin{pmatrix}
      \rho_s(\tau)\\
      \varphi_s(\tau)
    \end{pmatrix},
    \qquad 1\leq s\leq p-1.
  \end{gather*}

  It follows that under $\repA$, we have the decomposition
  \begin{equation*}
    \cZ_{\mathrm{cft}}
    =\underbrace{\oC\oplus\dots\oplus\oC}_{p+1}
    \oplus\underbrace{\oC^2\oplus\dots\oplus\oC^2}_{p-1}
  \end{equation*}
  (where $\oC$ is the trivial representation) and under $\repSw$, the
  decomposition
  \begin{equation*}
    \cZ_{\mathrm{cft}}
    =\mathcal{R}_{p+1}\oplus\mathcal{R}'_{p-1}\oplus\mathcal{R}''_{p-1}.
  \end{equation*}
  It is now straightforward to verify that $\repSw$ and $\repA$
  satisfy the required relations.
\end{proof}

\subsubsection{Remarks}\label{rem-to-1-thm}\mbox{}

\begin{enumerate}
  
\item Up to some simple multipliers, $\repA$ is just the inverse
  matrix automorphy factor in~\cite{[FHST]} and the restriction of
  $\repSw$ to $\mathcal{R}_{p+1}\oplus\mathcal{R}''_{p-1}$ is the
  $\SLiiZ$-representation in~\cite{[FHST]} that leads to the fusion
  algebra~\eqref{the-fusion} via a nonsemisimple generalization of the
  Verlinde formula.
  
\item $\mathcal{R}_{p-1}$ is the $\SLiiZ$-rep\-resentation realized in
  the $\hSL2_{p-2}$ minimal model~\cite{[Kac],[JFuchs]}.
  
  
\end{enumerate}

In Sec.~\ref{sec:SLiiZ-restr}, the structure described
in~\bref{thm:R-decomp} is established for the $\SLiiZ$-representation
on the quantum group center.

\section{$\UresSL2$: representations and the Grothendieck
  ring}\label{sec:all-Usl2} The version of the quantum $s\ell(2)$ that
is Kazhdan--Lusztig-dual to the $(1,p)$ conformal field theory model
is the restricted quantum group $\UresSL2$ at $\q$ the primitive
$2p$th root of unity.  We introduce it in~\bref{sec:Cas}, consider its
representations in~\bref{sec:repr}, and find its Grothendieck ring
in~\bref{sec:Grring}.

\subsection{The restricted quantum group $\UresSL2$}\label{sec:Cas}
The Hopf algebra $\UresSL2$ (henceforth, at $\q=e^{\frac{i\pi}{p}}$)
is generated by $E$, $F$, and $K$ with the relations
\begin{equation*}
  E^{p}=F^{p}=0,\quad K^{2p}=\one
\end{equation*}
and the Hopf-algebra structure given by
\begin{gather*}
  KEK^{-1}=\q^2E,\quad
  KFK^{-1}=\q^{-2}F,\\
  [E,F]=\ffrac{K-K^{-1}}{\q-\q^{-1}},\\
  \Delta(E)=\one\otimes E+E\otimes K,\quad
  \Delta(F)=K^{-1}\otimes F+F\otimes\one,\quad
  \Delta(K)=K\otimes K,\\
  \epsilon(E)=\epsilon(F)=0,\quad\epsilon(K)=1,\\
  S(E)=-EK^{-1},\quad  S(F)=-KF,\quad S(K)=K^{-1}.
\end{gather*}

The elements of the PBW-basis of $\UresSL2$ are enumerated as
$E^i\,K^j\,F^\ell$ with $0\leq i\leq p-1$, $0\leq j\leq 2p-1$,
$0\leq\ell\leq p-1$, and its dimension is therefore~$2p^3$.

\subsubsection{}It follows (e.g., by induction) that
\begin{multline}\label{Delta-formula}
  \Delta(F^m E^n K^j)
  =\sum_{r=0}^m\sum_{s=0}^n\,\q^{2(n-s)(r-m)+r(m - r) + s(n - s)}
  \qbin{m}{r}\qbin{n}{s}\\*
  {}\times
  F^r E^{n-s} K^{r-m+j}\tensor F^{m-r} E^s K^{n-s+j}.
\end{multline}

\subsubsection{The (co)integral and the comodulus}\label{sec:int}
For $\UresSL2$, the right integral and the left--right
cointegral (see the definitions in~\bref{app:int}) are given by
\begin{gather*}
  \rint(F^i E^m K^n)
  =
  \ffrac{1}{\zeta}\,\delta_{i,p-1}\delta_{m,p-1}\delta_{n, p+1}
\end{gather*}
and
\begin{gather}\label{coint}
  \coint=\zeta\,F^{p-1}E^{p-1}\sum_{j=0}^{2p-1}K^j,
\end{gather}
where we choose the normalization as
\begin{equation*}
  \zeta=\sqrt{\ffrac{p}{2}}\,\ffrac{1}{([p-1]!)^2}
\end{equation*}
for future convenience.

Next, simple calculation shows that the comodulus for $\UresSL2$
(see~\bref{app:int}) is $\comodul=K^2$.  This allows us to find the
balancing element using~\eqref{bal-comod}.  There are \textit{four}
possibilities for the square root of~$\comodul$, two of which are
group-like, and we choose
\begin{equation}\label{M-balance}
  \balance=K^{p+1}.
\end{equation}
This choice determines a ribbon element for~$\UresSL2$, and hence a
particular version of the $\SLiiZ$-action on the quantum group studied
below.

The balancing element~\eqref{M-balance} allows constructing the
``canonical'' $q$-characters of~$\UresSL2$-representations
(see~\bref{app:qCh}).

\subsubsection{The Casimir element}\label{sec:Casimir} Let $\cZ$
denote the center of $\UresSL2$.  It contains the element
\begin{equation}\label{eq:casimir}
  \cas=EF+\ffrac{\q^{-1}K+\q K^{-1}}{(\q-\q^{-1})^2}=
  FE+\ffrac{\q K+\q^{-1}K^{-1}}{(\q-\q^{-1})^2},
\end{equation}
called the Casimir element.  It satisfies the minimal polynomial
relation 
\begin{equation}\label{Cas-relation}
  \Psi_{2p}(\cas)=0,
\end{equation}
where 
\begin{equation*}
  \Psi_{2p}(x) =
  (x-\beta_0)\,(x-\beta_p)\prod_{j=1}^{p-1}(x-\beta_j)^2, \quad
  \beta_j=\ffrac{\q^j+\q^{-j}}{(\q-\q^{-1})^2}.
\end{equation*}
A proof of~\eqref{Cas-relation} is given in~\bref{fusion-center} below
as a spin-off of the technology developed for the Grothendieck ring
(we do not need~\eqref{Cas-relation} before that).

It follows from the definition of~$\UresSL2$ that $K^p\,{\in}\,\cZ$.
In fact, $K^p$ is in the $2p$-dimensional subalgebra in $\cZ$
generated by~$\cas$ because of the identity
\begin{equation}\label{Kp-Cas}
  K^{p} = \half
  \sum_{r=0}^{\floor{\frac{p}{2}}}
  \ffrac{p}{p-r}\,
  \mbox{\footnotesize$\displaystyle\binom{p-r}{r}$}
  (-1)^{1 - r}\,\hat\cas^{p - 2r},
\end{equation}
where we set
\begin{equation*}
  \hat\cas=(\q-\q^{-1})^2\cas.
\end{equation*}

\subsection{$\UresSL2$-representations}
\label{sec:repr} The $\UresSL2$-representation theory 
at $\q=e^{\frac{i\pi}{p}}$ is not difficult to describe (also
see~\cite{[RT],[GL],[JMT]}).  There turn out to be just $2p$
irreducible representations.  In what follows, we also need Verma
modules (all of which except two are extensions of a pair of
irreducible representations) and projective modules (which are further
extensions).  The category of all finite-dimensional
$\UresSL2$-representations at the primitive $2p$th root of unity is
fully described in~\cite{[FGST2]}.

\subsubsection{Irreducible representations}
\label{subsec:irrep}
The irreducible $\UresSL2$-representations $\repX^{\alpha}(s)$ are
labeled by $\alpha=\pm1$ and $1\leq s\leq p$.  The module
$\repX^{\pm}(s)$ is linearly spanned by elements $\vectv{s,n}^{\pm}$,
$0\leq n\leq s-1$, where $\vectv{s,0}^{\pm}$ is the highest-weight
vector and the $\UresSL2$-action is given~by
\begin{align*}
  K \vectv{ s, n}^{\pm} &=
  \pm \q^{s - 1 - 2n} \vectv{ s, n}^{\pm},\\
  E \vectv{ s, n}^{\pm} &=
  \pm [n][s - n]\vectv{ s, n - 1}^{\pm},\\
  F \vectv{ s, n}^{\pm} &= \vectv{ s, n + 1}^{\pm},
\end{align*}
where we set $\vectv{ s, s}^{\pm}=\vectv{ s, -1}^{\pm}=0$.  \ 
$\repX^{+}(1)$ is the trivial module.

For later use, we list the weights occurring in the
module~$\repLambda(s)$, i.e., the eigenvalues that $K$ has on vectors
in~$\repLambda(s)$,
\begin{equation}\label{eq:K-eigens-Lambda}
  \q^{-s+1},\q^{-s+3},\dots,\q^{s-1},
\end{equation}
and in the module~$\repPi(p-s)$,
\begin{equation}\label{eq:K-eigens-Pi}
  \q^{s+1},\q^{s+3},\dots,\q^{2p-s-1}.
\end{equation}

We also note the dimensions and quantum dimensions
(see~\bref{app:qCh}) $\dim\repX^{\alpha}(s)=s$ and
$\qdim\repX^{\alpha}(s)=\alpha^{p-1}(-1)^{s-1}[s]$.  It follows that
$\qdim\repX^{\alpha}(s)=-\qdim\repX^{-\alpha}(p-s)$ and
$\qdim\repX^{\alpha}(p)=0$.

\subsubsection{Verma modules}\label{subsec:verma-mod} There are $2p$
Verma modules $\Verma^{\pm}(s)$, $1\leq s\leq p$.  First, these are
the two Steinberg modules
\begin{equation*}
  \Verma^{\pm}(p)=\repX^{\pm}(p).
\end{equation*}
Next, for each $s=1,\dots,p-1$ and $\alpha=\pm1$, the Verma module
$\Verma^{\alpha}(s)$ is explicitly described in~\bref{verma-mod-base}
as an extension $0\to\repX^{-\alpha}(p-s)\to
\Verma^{\alpha}(s)\to\repX^{\alpha}(s)\to0$; for consistency with more
complicated extensions considered below, we represent it as
\begin{equation*}
  \overset{\repX^{\alpha}(s)}{\bullet}{}\longrightarrow{}
  \overset{\!\!\!\repX^{-\alpha}(p-s)\!\!\!}{\bullet},
\end{equation*}
with the convention that the arrow is directed to a
\textit{sub}module.  We note that $\dim\Verma^{\alpha}(s)=p$ and
$\qdim\Verma^{\alpha}(s)=0$ (negligible modules~\cite{[Fink]}).

\subsubsection{Projective modules}\label{subsec:proj-mod}
For $s=1,\dots,p-1$, there are nontrivial extensions yielding the
projective modules $\modL(s)$ and $\modP(s)$,
\begin{gather*}
  0\to\Verma^{-}(p-s)\to \modL(s)\to\Verma^{+}(s)\to0,\\
  0\to\Verma^{+}(p-s)\to \modP(s)\to\Verma^{-}(s)\to0.
\end{gather*}
Their structure can be schematically depicted~as
\begin{equation}\label{schem-proj}
  \xymatrix@=12pt{
    &&\stackrel{\repX^{\alpha}(s)}{\bullet}
    \ar@/^/[dl]
    \ar@/_/[dr]
    &\\
    &\stackrel{\repX^{-\alpha}(p{-}s)}{\bullet}\ar@/^/[dr]
    &
    &\stackrel{\repX^{-\alpha}(p{-}s)}{\bullet}\ar@/_/[dl]
    \\
    &&\stackrel{\repX^{\alpha}(s)}{\bullet}&
  }
\end{equation}
It follows that $\dim\modL(s)=\dim\modP(s)=2p$ and
$\qdim\modL(s)=\qdim\modP(s)=0$.  The bases and the action of
$\UresSL2$ in $\modL(s)$ and $\modP(s)$ are described
in~\bref{module-L} and~\bref{module-P}.

\subsection{The Grothendieck ring}\label{sec:Grring}  We next find the
Grothendieck ring of $\UresSL2$.

\begin{thm}\label{thm:Gr-ring}
  Multiplication in the $\UresSL2$ Grothendieck ring $\Grring_{2p}$ is
  given by
  \begin{equation*}
    \repX^{\alpha}(s)\,\repX^{\alpha'}(s')
    =\smash[b]{\sum_{\substack{s''=|s - s'| + 1\\
          \mathrm{step}=2}}^{s + s' - 1}}
    \widetilde\repX^{\alpha\alpha'}(s''),
  \end{equation*}
  where
  \begin{equation*}
    \widetilde\repX^{\alpha}(s)
    =
    \begin{cases}
      \repX^{\alpha}(s),&1\leq s\leq p,\\
      \repX^{\alpha}( 2p - s) + 2\repX^{-\alpha}( s - p),
      & p + 1 \leq s \leq 2p - 1.
    \end{cases}
  \end{equation*}
\end{thm}

To prove this, we use (i)~a property of the tensor products of any
representation with a Verma module, (ii)~an explicit evaluation of the
tensor product of any irreducible representation with a
two-dimensional one, and (iii)~the observation that the information
gained in (i) and (ii) suffices for finding the entire Grothendieck
ring.

We first of all note that the trivial representation $\repLambda(1)$
is the unit in the Grothen\-dieck ring and, obviously,
\begin{equation*}
  \repX^{\alpha}(s)\,\repPi(1)=\repX^{-\alpha}(s)
\end{equation*}
for all $s=1,\dots,p$ and $\alpha=\pm1$.  Moreover,
\begin{equation*}
  \repX^{\alpha}(s)\,\repPi(s')
  =\repX^{\alpha}(s)\,\repLambda(s')\,\repPi(1)
  =\repX^{-\alpha}(s)\,\repLambda(s'),
\end{equation*}
and it therefore suffices to find all the products
$\repX^{\alpha}(s)\,\repLambda(s')$ and, furthermore, just the
products $\repX^{+}(s)\,\repX^+(s')$.

\subsubsection{Products with Verma modules} In the Grothendieck ring,
the Verma module $\Verma^{\alpha}(s)$ (with $1\leq s\leq p-1$) is
indistinguishable from $\Verma^{-\alpha}(p-s)$, and we choose to
consider only the $p+1$ Verma modules~$\Verma_a$, $a=0,1,\dots,p$,
given by
\begin{equation}\label{the-Verma}
  \Verma_{0}=\Verma^{-}(p),\qquad
  \Verma_{a}=\Verma^{+}(a),\quad 1\leq a\leq p-1,
  \qquad
  \Verma_{p}=\Verma^{+}(p).
\end{equation}
Their highest weights $\q^{a-1}$ coincide with the respective highest
weights of $\repPi(p)$, $\repLambda(a)$, $\repLambda(p)$.

Taking the tensor product of a Verma module $\Verma_a$ and an
irreducible representation gives a module that is filtered by Verma
modules.  In the Grothendieck ring, this tensor product therefore
evaluates as a sum of Verma modules; moreover, the Verma modules that
occur in this sum are known, their highest weights being given by
$\q^{\varepsilon_a + \varepsilon_i}$, where $\q^{\varepsilon_a}$ is
the highest weight of $\Verma_a$ and $\q^{\varepsilon_i}$ are the
weights of vectors in the irreducible representation.
With~\eqref{eq:K-eigens-Lambda}, this readily gives the
Grothendieck-ring multiplication
\begin{gather}\label{decompose-Verma}
  \Verma_a\,\repLambda(s') =\sum_{\substack{s''=a-s'+1\\
      \text{step}=2}}^{a+s'-1} \Verma_{s''},
\end{gather}
where we set $\Verma_{s''}= \Verma_{-s''}$ for $s''<0$ and
$\Verma_{p+s''}= \Verma_{p-s''}$ for $0<s''<p$.

\begin{lemma}
  For $2\leq s\leq p-1$, we have
  \begin{equation*}
    \repX^{\alpha}(s)\,\repLambda(2)
    = \repX^{\alpha}(s - 1) + \repX^{\alpha}(s + 1).
  \end{equation*}
\end{lemma}
\begin{proof}   
  Let $e_{k}=\vectv{ s, k}^{\alpha}$ for $0\leq k\leq s-1$ and
  $\{f_{0}=\vectv{ 2, 0}^{+}, f_{1}=\vectv{ 2, 1}^{+}\}$ be the
  respective bases in $\repX^{\alpha}(s)$ and in $\repLambda(2)$.
  Under the action of $F$, the highest-weight vector $e_{0}\otimes
  f_{0}$ with the weight $\alpha \q^{(s+1)-1}$ generates the module
  $\repX^{\alpha}(s+1)$.  The vector $e'_{0}=e_{1}\otimes f_{0} -
  \alpha \q[s-1]e_{0}\otimes f_{1}$ satisfies the relations
  \begin{equation*}
    E e'_{0}=0, \quad K e'_{0}=\alpha \q^{(s-1)-1}e'_{0}.
  \end{equation*}
  Under the action of $F$, it generates the module
  $\repX^{\alpha}(s-1)$.
\end{proof}

As regards the product $\repX^{\alpha}(p)\,\repLambda(2)$, we already
know it from \eqref{decompose-Verma} because $\repX^{\alpha}(p)$ is a
Verma module: with the two relevant Verma modules replaced by the sum
of the corresponding irreducible representations, the resulting four
terms can be written as
\begin{equation*}
  \repX^{\alpha}(p)\,\repLambda(2)
  = 2\repX^{\alpha}(p-1)+2\repX^{-\alpha}(1).
\end{equation*}

As we have noted, the products $\repX^{\alpha}(s)\,\repPi(2)$ are
given by the above formulas with the reversed ``$\alpha$'' signs in
the right-hand sides.

\subsubsection{} We next evaluate the products
$\repX^{\alpha}(s)\,\repLambda(3)$ as
\begin{equation*}
  \repX^{\alpha}(s)\,\repLambda(3)
  =\repX^{\alpha}(s){}\,{}
  \bigl(\repLambda(2)\,\repLambda(2)-\repLambda(1)\bigr),
\end{equation*}
where the products with $\repLambda(2)$ are already known.  By
induction on $s'$, this allows finding all the products
$\repX^{\alpha}(s)\,\repLambda(s')$ as
\begin{multline}\label{rewrite_v2}
  \repX^{\alpha}(s)\,\repX^{+}(s')=
  \smash[b]{\sum_{\substack{s''=|s - s'| + 1\\s''\neq p,\;
        \mathrm{step}=2}}^{p - 1 - |p - s - s'|}}
  \repX^{\alpha}(s'') +
  \delta_{p, s, s'}\repX^{\alpha}(p)\\*
  {}+ \sum_{\substack{s''= 2p - s - s' + 1\\ 
      \mathrm{step}=2}}^{p - 1}
  (2\repX^{\alpha}(s'') + 2\repX^{-\alpha}(p - s'')),
\end{multline}
where $\delta_{p,s,s'}$ is equal to~$1$ if $p-s-s'+1\leq 0$ and
$p-s-s'+1\equiv0\;\mathrm{mod}\;2$, and is~$0$ otherwise.

The statement in~\bref{thm:Gr-ring} is a mere rewriting
of~\eqref{rewrite_v2}, taken together with the relations
$\repX^{\alpha}(s)\,\repPi(s') =\repX^{-\alpha}(s)\,\repLambda(s')$.
It shows that the $\UresSL2$ Groth\-end\-ieck ring is the
$(1,p)$-model fusion algebra derived in~\cite{[FHST]}.
This concludes the proof of~\bref{thm:Gr-ring}.

\begin{cor}\label{cor:quotient}  
  The $\UresSL2$ Grothendieck ring contains the ideal~$\Videal_{p+1}$
  of Verma modules generated by
  \begin{equation*}
    \begin{aligned}
      &\repX^{+}(p-s)+\repX^{-}(s),\quad 1\leq s\leq p-1,\\
      &\repX^{+}(p),\quad \repX^{-}(p).
    \end{aligned}
  \end{equation*}
  The quotient~$\Grring_{2p}/\Videal_{p+1}$ is a \textit{fusion}
  algebra with the basis $\overline\repX(s)$, $1\leq s\leq p-1$
  \textup{(}the canonical images of the corresponding
  $\repX^{+}(s)$\textup{)} and multiplication
  \begin{equation*}
    \overline\repX(s)\,\overline\repX(s')
    = \sum_{\substack{s''=|s - s'| + 1\\
        \mathrm{step}=2}}^{p - 1 - |p - s - s'|}\overline\repX(s''),
    \quad s,s'=1,\dots,p-1.
  \end{equation*}
  This is a \emph{semisimple} fusion algebra, which coincides with the
  fusion of the unitary $\hSL2$ representations of level~$p-2$.
\end{cor}

\begin{cor}\label{cor:generated}
  The $\UresSL2$ Grothendieck ring $\Grring_{2p}$ is generated by
  $\repX^{+}(2)$.
\end{cor}
This easily follows from Theorem~\ref{thm:Gr-ring}; therefore,
$\Grring_{2p}$ can be identified with a quotient of the polynomial
ring~$\oC[x]$.  Let~$\cheb_s(x)$ denote the Chebyshev polynomials of
the second kind
\begin{equation}\label{eq:chebyshev-sin}
  \cheb_s(2\cos t)=\mfrac{\sin s t}{\sin t}.
\end{equation}
The lower such polynomials are $\cheb_0(x)=0$, $\cheb_1(x)=1$,
$\cheb_2(x)=x$, and $\cheb_3(x)=x^2-1$.

\begin{prop}\label{prop:quotient}
  The $\UresSL2$ Grothendieck ring is the quotient of the polynomial
  ring $\oC[x]$ over the ideal generated by the polynomial
  \begin{equation}\label{Cas-eq}
    \hat\Psi_{2p}(x)=
    \cheb_{2p+1}(x)-\cheb_{2p-1}(x)-2.
  \end{equation}
  Moreover, let
  \begin{equation}\label{basis-P}
    P_s(x)=
    \begin{cases}
      \cheb_s(x),& 1\leq s\leq p,\\
      \half\cheb_s(x)-\half\cheb_{2p-s}(x),& p+1\leq s\leq 2p.
    \end{cases}
  \end{equation}
  Under the quotient mapping, the image of each polynomial $P_s$
  coincides with $\repX^+(s)$ for $1\leq s\leq p$ and with
  $\repX^-(s-p)$ for $p+1\leq s\leq2p$.
\end{prop}
\begin{proof}
  It follows from~\bref{thm:Gr-ring} that 
  \begin{align}
  \repX^+(2)\,\repX^\pm(1)&=\repX^\pm(2),\label{rel1}\\
  \repX^+(2)\,\repX^\pm(s)&=\repX^\pm(s-1)+\repX^\pm(s+1),\quad2\leq
  s\leq p-1,\\
  \repX^+(2)\,\repX^+(p)&=2\repX^+(p-1)+2\repX^-(1),\label{rel4}\\ 
  \repX^+(2)\,\repX^-(p)&=2\repX^-(p-1)+2\repX^+(1).\label{rel5}
  \end{align}
  We recall that the Chebyshev polynomials of the second kind satisfy
  (and are determined by) the recursive relation
  \begin{equation}\label{cheb-rec}
    x\cheb_s(x)=\cheb_{s-1}(x)+\cheb_{s+1}(x),
    \quad s \geq 2,
  \end{equation}
  with the initial data $\cheb_1(x)=1$, $\cheb_2(x)=x$.  {}From
  \eqref{cheb-rec}, we then obtain that polynomials~\eqref{basis-P}
  satisfy relations \eqref{rel1}--\eqref{rel4} after the
  identifications $P_s\to\repX^+(s)$ for $1\leq s\leq p$ and
  $P_s\to\repX^-(s-p)$ for $p+1\leq s\leq2p$.  Then, for
  Eq.~\eqref{rel5} to be satisfied, we must impose the relation
  $xP_{2p}(x)\equiv 2P_{2p-1}(x)+2P_1(x)$; this shows that the
  Grothendieck ring is the quotient of $\oC[x]$ over the ideal
  generated by polynomial~\eqref{Cas-eq}.
 \end{proof}

\medskip

\begin{prop}\label{prop:factor}\mbox{}
  The polynomial $\hat\Psi_{2p}(x)$ can be factored
  as 
  \begin{equation*}
    \qquad\hat\Psi_{2p}(x) =
    (x-\hat\beta_0)\,(x-\hat\beta_p)\prod_{j=1}^{p-1}(x-\hat\beta_j)^2,
    \quad \hat\beta_j=\q^j+\q^{-j}=2\cos\ffrac{\pi
      j}{p}.
  \end{equation*}
\end{prop}
This is verified by direct calculation using the representation
\begin{equation*}
  \hat\Psi_{2p}(2\cos
  t)=2(\cos(2 p t) - 1),
\end{equation*}
which follows from~\eqref{eq:chebyshev-sin}.  We note that
$\hat\beta_j\neq\hat\beta_{j'}$ for $0\leq j\neq j'\leq p$.

\section{$\UresSL2$: factorizable and ribbon Hopf algebra structures
  \hbox{and the center}}\label{sec:new} The restricted quantum group
$\UresSL2$ is not quasitriangular~\cite{[ChP]}; however, it admits a
Drinfeld mapping, and hence there exists a homomorphic image
$\Drinalg_{2p}$ of the Grothendieck ring in the center.
In~\bref{sec:from}, we first identify $\UresSL2$ as a subalgebra in a
quotient of a Drinfeld double.  We then obtain the $M$-matrix
in~\bref{sec:M-matrix}, characterize the subalgebra
$\Drinalg_{2p}\subset\cZ$ in~\bref{fusion-center}, and find the center
$\cZ$ of $\UresSL2$ at $\q=e^{\frac{i\pi}{p}}$ in~\bref{the-center}.
Furthermore, we give some explicit results for the Radford mapping
for~$\UresSL2$ in~\bref{sec:Radford-SL2} and we find a ribbon element
for~$\UresSL2$ in~\bref{SL2-ribbon}.

\subsection{$\UresSL2$ from the double}\label{sec:from}
The Hopf algebra $\UresSL2$ is not quasitriangular, but it can be
realized as a Hopf subalgebra of a quasitriangular Hopf algebra $\bar
D$ (which is in turn a quotient of a Drinfeld double).  The $M$-matrix
(see~\bref{app:M}) for $\bar D$ is in fact an element of
$\UresSL2\tensor\UresSL2$, and hence $\UresSL2$ can be thought of as a
factorizable Hopf algebra, even though relation~\eqref{M-Delta} required
of an $M$-matrix is satisfied not in $\UresSL2$ but in~$\bar D$ (but
on the other hand, \eqref{M-factorizable} holds only with $\pbwd_I$ and
$\pbwdd_I$ being bases in~$\UresSL2$).

The Hopf algebra $\bar D$ is generated by $\dE$, $\dF$, and $\dK$ with
the relations
\begin{gather*}
  \dK\dE\dK^{-1}=\q\dE,\quad\dK\dF\dK^{-1}=\q^{-1}\dF,
  \quad[\dE,\dF]=\ffrac{\dK^2-\dK^{-2}}{\q-\q^{-1}},
  \\
  \dE^p=0,\quad\dF^p=0,\quad\dK^{4p}=\one, 
  \\
  \epsilon(\dE)=0,\quad\epsilon(\dF)=0,\quad
  \epsilon(\dK)=1,
  \\
  \Delta(\dE)=\one\otimes\dE+\dE\otimes\dK^2,\quad
  \Delta(\dF)=\dK^{-2}\tensor\dF+\dF\tensor\one,\quad
  \Delta(\dK)=\dK\otimes\dK, 
  \\
  S(\dE)=-\dE\dK^{-2},\quad S(\dF)=-\dK^{2}\dF,\quad
  S(\dK)=\dK^{-1}.
\end{gather*}

A Hopf algebra embedding $\UresSL2\to\bar D$ is given by
\begin{equation*}
  E\mapsto\dE,\quad F\mapsto\dF,\quad
  K\mapsto\dK^2.
\end{equation*}
In what follows, we often do not distinguish between $E$ and $\dE$,
$F$ and $\dF$, and $K$ and~$\dK^2$.

\begin{thm}\label{thm-bar-one}
  $\bar D$ is a ribbon quasitriangular Hopf algebra, with the universal
  $R$-matrix
  \begin{equation}\label{bar-R}
    \bar R =\ffrac{1}{4p}\sum_{m=0}^{p-1}\sum_{n,j=0}^{4p-1}
    \ffrac{(\q-\q^{-1})^m}{[m]!}\,\q^{m(m-1)/2+m(n-j)-nj/2}
    \dE^m\dK^{n}\otimes\dF^m\dK^{j}
  \end{equation}
  and the ribbon element
  \begin{equation}\label{rib-bar-D} \ribbon
    =\ffrac{1-i}{2\sqrt{p}}\sum_{m=0}^{p-1}\sum_{j=0}^{2p-1}
    \ffrac{(\q-\q^{-1})^m}{[m]!}\,
    \q^{-\frac{m}{2}+mj+\half(j+p+1)^2}\dF^m\dE^m\dK^{2j}.
  \end{equation}
\end{thm}
\begin{proof}
  Equation~\eqref{bar-R} follows from the realization of $\bar D$ as a
  quotient of the Drinfeld double $D(B)$ in~\bref{thm:double}.  The
  quotient is over the Hopf ideal generated by the central element
  $\ddK\dK-\one\in D(B)$.  It follows that $\bar D$ inherits a
  quasitriangular Hopf algebra structure from $D(B)$ and
  $R$-matrix~\eqref{bar-R} is the image of~\eqref{the-R} under the
  quotient mapping.

  Using $R$-matrix~\eqref{bar-R}, we calculate the canonical
  element~$\sqs$ (see~\eqref{canon-sqs}) as
  \begin{equation}\label{square-S}
    \sqs
    =\ffrac{1}{4p}\sum_{m=0}^{p-1}
    \sum_{n,r=0}^{4p-1}(-1)^m\ffrac{(\q-\q^{-1})^m}{[m]!}\,
    \q^{-m(m+3)/2-rn/2}\dF^m\dK^{-r}\dE^m\dK^n.
  \end{equation}
  We note that actually~$\sqs\,{\in}\,\UresSL2$. Indeed,
  \begin{multline*}
    \sqs=\ffrac{1}{4p}\sum_{m=0}^{p-1}\sum_{n,r=0}^{4p-1}
    (-1)^m\ffrac{(\q-\q^{-1})^m}{[m]!}\,
    \q^{-m(m+3)/2-rm-rn/2}\dF^m\dE^m\dK^{n-r}={}\\
    =\ffrac{1}{4p}\sum_{m=0}^{p-1}\sum_{j=0}^{2p-1}
    \Bigl(\sum_{r=0}^{4p-1}e^{-i\pi\frac{1}{2p}r(r+2m+2j)}\Bigr)
    (-1)^m\ffrac{(\q-\q^{-1})^m}{[m]!}\,
    \q^{-\half m(m+3)}\dF^m\dE^m\dK^{2j}\\
    + \ffrac{1}{4p}\sum_{m=0}^{p-1}\sum_{j=0}^{2p-1}
    \Bigl(\sum_{r=0}^{4p-1}e^{-i\pi\frac{1}{2p}r(r+2m+2j+1)}\Bigr)
    (-1)^m\ffrac{(\q-\q^{-1})^m}{[m]!}\, \q^{-\half
      m(m+3)}\dF^m\dE^m\dK^{2j+1}.
  \end{multline*}
  The second Gaussian sum vanishes,
  \begin{equation*}
    \sum_{r=0}^{4p-1}e^{-i\pi\frac{1}{2p}r(r+2m+2j+1)}=0.
  \end{equation*}
  To evaluate the first Gaussian sum, we make the substitution $r\to
  r-j-m$:
  \begin{multline*}
    \sqs=\ffrac{1}{4p}\sum_{m=0}^{p-1}\sum_{j=0}^{2p-1}
    \Bigl(\sum_{r=0}^{4p-1}e^{-i\pi\frac{1}{2p}r^2}\Bigr)
    (-1)^m\ffrac{(\q-\q^{-1})^m}{[m]!}\,
    \q^{-\half m(m+3)+\half(j+m)^2}\dF^m\dE^m\dK^{2j}\\
    =\ffrac{1}{4p}\sum_{m=0}^{p-1}\sum_{j=0}^{2p-1}
    \Bigl(\sum_{r=0}^{4p-1}e^{-i\pi\frac{1}{2p}r^2}\Bigr)
    \ffrac{(\q-\q^{-1})^m}{[m]!}\, \q^{-\half m+m(j-p-1)+\half
      j^2}\dF^m\dE^m\dK^{2j}.
  \end{multline*}
  Then evaluating
  \begin{equation*}
    \sum_{r=0}^{4p-1}e^{-i\pi\frac{1}{2p}r^2}=(1-i)2\sqrt{p},
  \end{equation*}
  we obtain
  \begin{equation*}
    \sqs=\ffrac{1-i}{2\sqrt{p}}\sum_{m=0}^{p-1}\sum_{j=0}^{2p-1}
    \ffrac{(\q-\q^{-1})^m}{[m]!}\,
    \q^{-\half m+mj+\half (j+p+1)^2}\dF^m\dE^m\dK^{2j+2p+2}.
  \end{equation*}
  
  We then find the ribbon element from relation~\eqref{balance-ribbon}
  using the balancing
  element~$\balance=\dK^{2p+2}$\label{page:balancing}
  from~\eqref{M-balance}, which gives~\eqref{rib-bar-D}.
\end{proof}

\subsection{The $M$-matrix for $\UresSL2$}\label{sec:M-matrix}
We next obtain the $M$-matrix (see~\bref{app:M}) for $\UresSL2$ from
the universal $R$-matrix for $\bar D$ in~\eqref{bar-R}.  Because
$\sqs\in\UresSL2$, it follows from~\eqref{Delta-u} that the $M$-matrix
for $\bar D$, \ $\bar M=\bar R_{21}\bar R_{12}$, actually lies in
$\UresSL2\tensor\UresSL2$, and does not therefore satisfy
condition~\eqref{M-factorizable} in~$\bar D$ (and hence $\bar D$ is not
factorizable).  But this \textit{is} an $M$-matrix for
$\UresSL2\subset\bar D$.  A simple calculation shows that $\bar
R_{21}\bar R_{12}$ is explicitly rewritten in terms of the
$\UresSL2$-generators as
\begin{multline}\label{bar-M}
  \bar M
  =\ffrac{1}{2p}
  \sum_{m=0}^{p-1}\sum_{n=0}^{p-1}
  \sum_{i=0}^{2p-1}\sum_{j=0}^{2p-1}
  \ffrac{(\q - \q^{-1})^{m + n}}{[m]! [n]!}\,
  \q^{m(m - 1)/2 + n(n - 1)/2}\\*
  \times \q^{- m^2 - m j + 2n j - 2n i - i j + m i} 
  F^{m} E^{n} K^{j}\tensor E^{m} F^{n} K^{i}.
\end{multline}

\subsection{Drinfeld mapping and the $(1,p)$ fusion in
  $\cZ(\UresSL2)$}\label{fusion-center} Given the $M$-matrix, we can
identify the $\UresSL2$ Grothendieck ring with its image in the center
using the homomorphism in~\bref{lemma:Dr-hom}.  We evaluate this
homomorphism on the preferred basis elements in the Grothendieck ring,
i.e., on the irreducible representations.  With the balancing element
for $\UresSL2$ in~\eqref{M-balance} and the $M$-matrix
in~\eqref{bar-M}, the mapping in~\bref{lemma:Dr-hom}~is
\begin{equation}\label{cchi-def}
  \begin{split}
    \Grring_{2p}&\to\cZ\\    
    \repX^{\pm}(s)&\mapsto\cchi^{\pm}(s)\equiv
    \drmap(\qTr_{\repX^{\pm}(s)})=
    (\Tr_{\repX^{\pm}(s)}\tensor\id)
    \bigl((K^{p-1}\tensor\one)\,
    \bar M 
    \bigr),
    \quad
    1\leq s\leq p.
  \end{split}
\end{equation}
Clearly, $\cchi^{+}(1)=\one$.  We let $\Drinalg_{2p}\subset\cZ$ denote
the image of the Grothendieck ring under this mapping.

\begin{prop}\label{prop-eval}
  For $s=1,\dots,p$ and $\alpha=\pm1$,
  \begin{multline}\label{the-cchi}
    \cchi^{\alpha}(s) = \alpha^{p+1}(-1)^{s+1}\sum_{n=0}^{s-1}
    \sum_{m=0}^{n}
    (\q-\q^{-1})^{2m} \q^{-(m+1)(m+s-1-2n)}\times{}\\*
    {}\times\qbin{s-n+m-1}{m}
    \qbin{n}{m} E^m F^m K^{s-1+\beta p - 2n + m},
  \end{multline}
  where we set $\beta=0$ if $\alpha=+1$ and $\beta=1$ if $\alpha=-1$.
  In particular, it follows that
  \begin{gather}\label{chi12}
    \cchi^{+}(2)=-\hat\cas\\
    \intertext{\textup{(}with $\hat\cas$ defined
      in~\bref{sec:Casimir}\textup{)} and}
    \label{minus-alpha}
    \cchi^{-\alpha}(s)=-(-1)^p\cchi^{\alpha}(s) K^p.
  \end{gather}
\end{prop}
\begin{proof}
  The proof of~\eqref{the-cchi} is a straightforward calculation based
  on the well-known identity (see, e.g.,~\cite{[ChP]})
  \begin{equation}\label{eq:EmFm-prod2}
    \prod_{s=0}^{r-1}
    \Bigl(\cas-\mfrac{\q^{2s+1}K+\q^{-2s-1}K^{-1}}{(\q-\q^{-1})^2}
    \Bigr)=
    F^rE^r,\qquad r<p,
  \end{equation}
  which readily implies that
  \begin{equation}\label{the-trace2}
    \Tr_{\repX^{\alpha}(s)} F^m E^m K^{a}
    =\alpha^{m+a}([m]!)^2\sum_{n=0}^{s-1}\q^{a(s-1-2n)}
    \qbin{s-n+m-1}{m}\qbin{n}{m}.
  \end{equation}
  Using this in~\eqref{cchi-def} gives~\eqref{the-cchi}.  For
  $\cchi^{+}(2)$, we then have
  \begin{multline*}
    \cchi^{+}(2)
    = -\sum_{n=0}^{1}
    \sum_{m=0}^{n}
    (\q-\q^{-1})^{2m} \q^{-(m+1)(m+1-2n)}
    \qbin{1-n+m}{m}\qbin{n}{m}
    E^m F^m K^{1 - 2n + m}=\\
    = -\q^{-1}K - \q K^{-1} - (\q - \q^{-1})^2 E F.
  \end{multline*}
\end{proof}

Combining~\bref{prop-eval} and~\bref{cor:generated}, we obtain
\begin{prop}\label{Dr-alg-Cas}
  $\Drinalg_{2p}$ coincides with the algebra generated by the Casimir
  element.
\end{prop}

The following corollary is now immediate in view
of~\bref{prop:quotient} and~\bref{prop:factor}.
\begin{cor}
  Relation~\eqref{Cas-relation} holds for the Casimir element.
\end{cor}

\begin{cor}
  Identity~\eqref{the-identity} holds.
\end{cor}
The derivation of~\eqref{the-identity} from the algebra of the
$\cchi^\alpha(s)$ is given in Appendix~\bref{app:derivation} in some
detail.  We note that although the left-hand side
of~\eqref{the-identity} is not manifestly symmetric in $s$ and $s'$,
the identity shows that it is.

\subsubsection{}\label{Verma-in-center}
In what follows, we keep the notation~$\Videal_{p+1}$ for the
Verma-module ideal (more precisely, for its image in the center)
generated by
\begin{equation}\label{verma-def}
  \begin{aligned}
    \vvarkappa(0)&=\cchi^{-}(p),\\
    \vvarkappa(s) &= \cchi^{+}(s) + \cchi^{-}(p-s),
    \quad 1\leq s\leq p-1,\\
    \vvarkappa(p)&=\cchi^{+}(p).
  \end{aligned}
\end{equation}
This ideal is the socle (annihilator of the radical)
of~$\Drinalg_{2p}$.

\subsection{The center of $\UresSL2$}\label{the-center} We now find
the center of $\UresSL2$ at the primitive $2p$th root of unity.  For
this, we use the isomorphism between the center and the algebra of
\textit{bimodule} endomorphisms of the regular representation.  The
results are in~\bref{prop-center} and~\bref{prop-center-explicit}.

\subsubsection{Decomposition of the regular representation}
The $2p^3$-dimensional regular representation of $\UresSL2$, viewed as
a free left module, decomposes into indecomposable projective modules,
each of which enters with the multiplicity given by the dimension of
its simple quotient:
\begin{equation*}
  \mathsf{Reg}=\bigoplus_{s=1}^{p-1} s\modL(s)
  \oplus
  \bigoplus_{s=1}^{p-1} s\modP(s)
  \oplus p\repLambda(p)\oplus p\repPi(p).
\end{equation*}

We now study the regular representation as a $\UresSL2$-bimodule.  In
what follows, $\boxtimes$ denotes the external tensor product.
\begin{prop}
  As a $\UresSL2$-bi\-module, the regular representation decomposes as
  \begin{equation*}
    \mathsf{Reg}=\smash[b]{\bigoplus_{s=0}^{p}\modQ(s)},
  \end{equation*}
  where
  \begin{enumerate}
  \item the bimodules
    \begin{equation*}
      \modQ(0)=\repPi(p)\boxtimes\repPi(p),\quad
      \modQ(p)=\repLambda(p)\boxtimes\repLambda(p)
    \end{equation*}
    are simple,
    
  \item the bimodules $\modQ(s)$, $1\leq s\leq p-1$, are indecomposable
    and admit the filtration
    \begin{equation}\label{filtr-Q}
      0\subset\repR_2(s)\subset\repR(s)\subset\modQ(s),
    \end{equation}
    where the structure of subquotients is given by
    \begin{gather}\label{prop:subquotient:Q/R}
      \modQ(s)/\repR(s)=\repLambda(s)\boxtimes\repLambda(s)
      \oplus\repPi(p-s)\boxtimes\repPi(p-s)
    \end{gather}
    and
    \begin{multline*}
      \repR(s)/\repR_2(s)=
      \repPi(p-s)\boxtimes\repLambda(s)\oplus\repPi(p-s)
      \boxtimes\repLambda(s)\\*
      {}\oplus\repLambda(s)\boxtimes\repPi(p-s)\oplus
      \repLambda(s)\boxtimes\repPi(p-s),
    \end{multline*}
    and where $\repR_2(s)$ is isomorphic to the quotient
    $\modQ(s)/\repR(s)$.
  \end{enumerate}
\end{prop}
The proof given below shows that $\repR(s)$ is in fact the Jacobson
radical of~$\modQ(s)$ and $\repR_2(s)=\repR(s)^2$, with
$\repR(s)\repR_2(s)=0$, and hence $\repR_2(s)$ is the socle
of~$\modQ(s)$.  For $s=1,\dots,p-1$, the left $\UresSL2$-action on
$\modQ(s)$ and the structure of subquotients can be visualized with
the aid of the diagram

\vspace*{-.8\baselineskip}

\begin{small}
  \begin{equation*}
    \mbox{}\kern-62pt
    \xymatrix@=16pt{%
      *{}&{\repLambda(s)\makebox[0pt][l]{${\boxtimes}\repLambda(s)$}}
      \ar[1,-1]
      \ar[1,1]
      &*{}&*{}&*{}
      *{}&{\repPi(p{-}s)\makebox[0pt][l]{${\boxtimes}\repPi(p{-}s)$}}
      \ar[1,-1]
      \ar[1,1]
      \\
      {\repPi(p{-}s)\makebox[0pt][l]{${\boxtimes}\repLambda(s)$}}
      \ar[1,1]
      &*{}&
      {\repPi(p{-}s)\makebox[0pt][l]{${\boxtimes}\repLambda(s)$}}
      \ar[1,-1]
      &*{\quad}
      &{\repLambda(s)\makebox[0pt][l]{${\boxtimes}\repPi(p{-}s)$}}
      \ar[1,1]
      &*{}
      &{\repLambda(s)\makebox[0pt][l]{${\boxtimes}\repPi(p{-}s)$}}
      \ar[1,-1]
      \\
      *{}&{\repLambda(s)\makebox[0pt][l]{${\boxtimes}\repLambda(s)$}}
      &*{}&*{}&*{}
      *{}&{\repPi(p{-}s)\makebox[0pt][l]{${\boxtimes}\repPi(p{-}s)$}}
    }
  \end{equation*}
\end{small}%
and the right action with
\begin{center}
  \includegraphics[bb=1.3in 9in 8in 10.2in, clip]{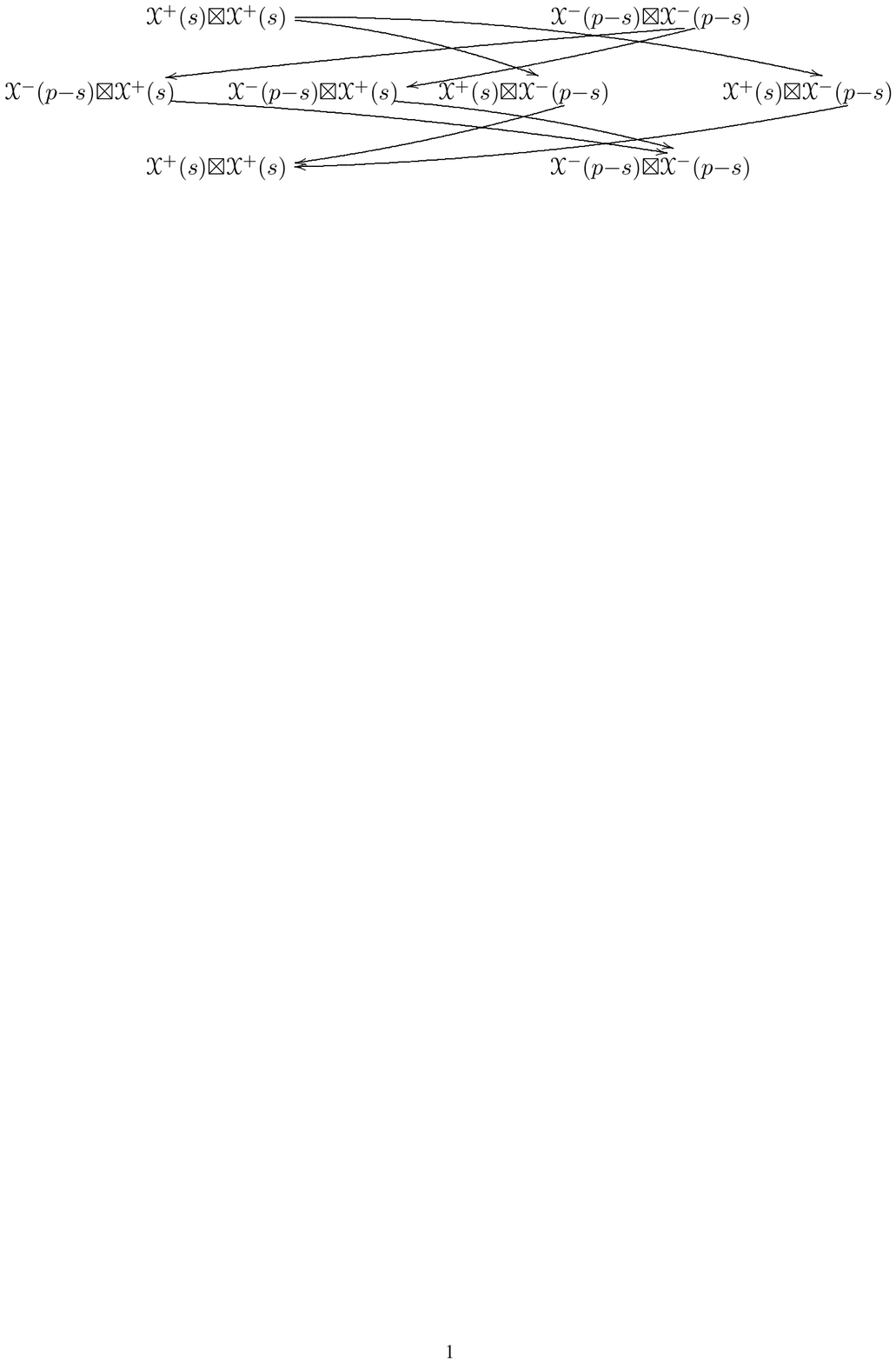}
\end{center}
The reader may find it convenient to look at these diagrams in reading
the proof below.
\begin{proof}
  First, the category $\lc$ of finite-dimensional left
  $\UresSL2$-modules has the decomposition~\cite{[FGST2]}
  \begin{equation}\label{decompose-cat}
    \lc=\bigoplus_{s=0}^{p}\lc(s),
  \end{equation}
  where each $\lc(s)$ is a full subcategory.  The full subcategories
  $\lc(0)$ and~$\lc(p)$ are semisimple and contain precisely one
  irreducible module each, $\repLambda(p)$ and~$\repPi(p)$
  respectively. Each $\lc(s)$, $1\,{\leq}\, s \,{\leq}\,p{-}1$,
  contains precisely two irreducible modules $\repLambda(s)$
  and~$\repPi(p{-}s)$, and we have the vector-space
  isomorphisms~\cite{[FGST2]}
  \begin{equation}\label{Ext-irr}
    \Ext(\repX^{\pm}(s),\repX^{\mp}(p-s))\cong \oC^2,
  \end{equation}
  where a basis in each $\oC^2$ can be chosen as the extensions
  corresponding to the Verma module $\Verma^{\pm}(s)$ and to the
  contragredient Verma module $\CVerma^{\pm}(s)$
  (see~\bref{verma-mod-base}).
  
  In view of~\eqref{decompose-cat}, the regular representation viewed
  as a $\UresSL2$-bimodule has the decomposition
  \begin{equation*}
    \mathsf{Reg}=\smash[t]{\bigoplus_{s=0}^{p}\modQ(s)}
  \end{equation*}
  into a direct sum of indecomposable two-sided ideals $\modQ(s)$.  We
  now study the structure of subquotients of $\modQ(s)$. Let
  $\repR(s)$ denote the Jacobson radical of $\modQ(s)$. By the
  Wedderburn--Artin theorem, the quotient $\modQ(s)/\repR(s)$ is a
  semisimple matrix algebra over~$\oC$,
  \begin{gather*}
    \modQ(s)/\repR(s)
    =\End(\repLambda(s))\oplus\End(\repPi(p\!-\!s)),\quad
    1\leq s\leq p-1,\\
    \modQ(0)=\End(\repPi(p)),\quad \modQ(p)=\End(\repLambda(p))
  \end{gather*}
  (where we note that $\repR(0)=\repR(p)=0$).  As a bimodule,
  $\modQ(s)/\repR(s)$ has the decomposition
  \begin{gather}\label{subquotient:Q/R}
    \modQ(s)/\repR(s)=\repLambda(s)\boxtimes\repLambda(s)
    \oplus\repPi(p\!-\!s)\boxtimes\repPi(p\!-\!s),\quad 1\leq s\leq p-1,\\
    \modQ(0)=\repPi(p)\boxtimes\repPi(p),\quad
    \modQ(p)=\repLambda(p)\boxtimes\repLambda(p).\notag
  \end{gather}
  
  For $1\leq s\leq p-1$, we now consider the quotient
  $\repR(s)/\repR_2(s)$, where we set~$\repR_2(s)=\repR(s)^2$.  For
  brevity, we write $\repR\equiv\repR(s)$, $\modQ\equiv\modQ(s)$,
  $\repLambda\equiv\repLambda(s)$ and $\repPi\equiv\repPi(p-s)$,
  $\Verma^{+}\equiv\Verma^{+}(s)$, $\Verma^{-}\equiv\Verma^{-}(p-s)$,
  and similarly for the contragredient Verma modules $\CVerma^{\pm}$.
  In view of \eqref{Ext-irr}, there are the natural bimodule
  homomorphisms
  \begin{equation*}
    \modQ\xrightarrow{\pi^{\pm}}\End(\Verma^{\pm}), \quad
    \modQ\xrightarrow{\bar{\pi}^{\pm}}\End(\CVerma^{\pm}).
  \end{equation*}
  The image of $\pi^+$ has the structure of the lower-triangular
  matrix
  \begin{equation*}
    \im(\pi^+)=
    \begin{pmatrix}
      \repLambda\boxtimes\repLambda & 0\\
      \repLambda\boxtimes\repPi & \repPi\boxtimes\repPi
    \end{pmatrix}
  \end{equation*}
  Clearly, the radical of $\im(\pi^+)$ is the bimodule
  $\repLambda\boxtimes\repPi$.  It follows that
  $\pi^+(\repR)=\repLambda\boxtimes\repPi$ and the bimodule
  $\repLambda\boxtimes\repPi$ is a subquotient of $\repR$.  In a
  similar way, we obtain that $\pi^-(\repR)=\repPi\boxtimes\repLambda$
  and $\bar{\pi}^{\pm}(\repR)=\repX^{\pm}\boxtimes\repX^{\mp}$.
  Therefore, we have the inclusion
  \begin{equation}\label{subquotient:R/RR2}    
    \repR/\repR^2\supset
    \repPi\boxtimes\repLambda\oplus\repPi\boxtimes\repLambda
    \oplus\repLambda\boxtimes\repPi\oplus\repLambda\boxtimes\repPi.
  \end{equation}
  
  Next, the Radford mapping $\radmap:\mathsf{Reg}^*\to \mathsf{Reg}$
  (see~\bref{sec:radford-all}) establishes a bimodule isomorphism
  between $\mathsf{Reg}^{*}$ and $\mathsf{Reg}$, and therefore the
  socle of~$\modQ$ is isomorphic to~$\modQ/\repR$.  This suffices for
  finishing the proof: by counting the dimensions of the subquotients
  given in~\eqref{subquotient:Q/R} and~\eqref{subquotient:R/RR2}, and
  the dimension of the socle of $\modQ$, we obtain the statement of
  the proposition.
\end{proof}

\subsubsection{Bimodule homomorphisms and the center} To find the
center of $\UresSL2$, we consider bimodule endomorphisms of the
regular representation; such endomorphisms are in a $1:1$
correspondence with elements in the center.  Clearly,
\begin{equation*}
  \End\bigl(\mathsf{Reg}\bigr)
  =\bigoplus_{s=0}^{p}\End\bigl(\modQ(s)\bigr).
\end{equation*}
For each $\modQ(s)$, $0\leq s\leq p$, there is a bimodule endomorphism
$\idem_s:\mathsf{Reg}\to\mathsf{Reg}$ that acts as identity on
$\modQ(s)$ and is zero on~$\modQ(s')$ with $s'\neq s$.  These
endomorphisms give rise to $p+1$ primitive idempotents in the center
of~$\UresSL2$.

Next, for each $\modQ(s)$ with $1\leq s\leq p-1$, there is a
homomorphism $\nilp^+_s:\modQ(s)\to\modQ(s)$ (defined up to a nonzero
factor) whose kernel, as a linear space, is given by
$\repR(s)\oplus\repPi(p-s)\boxtimes\repPi(p-s)$ (see~\eqref{filtr-Q});
in other words, $\nilp^+_s$ sends the quotient
$\repLambda(s)\boxtimes\repLambda(s)$ into the subbimodule
$\repLambda(s)\boxtimes\repLambda(s)$ at the bottom of $\modQ(s)$ and
is zero on~$\modQ(s')$ with $s'\neq s$.  Similarly, for each
$s=1,\dots,p-1$, there is a central element associated with the
homomorphism $\nilp^-_{s}:\modQ(s)\to\modQ(s)$ with the kernel
$\repR(s)\oplus\repLambda(s)\boxtimes\repLambda(s)$, i.e., the
homomorphism sending the quotient
$\repPi(p\,{-}\,s)\boxtimes\repPi(p-s)$ into the subbimodule
$\repPi(p-s)\boxtimes\repPi(p-s)$ (and acting by zero on~$\modQ(s')$
with $s'\neq s$).  In total, there are $2(p-1)$ elements
$\nilp^\pm_s$, $1\leq s\leq p-1$, which are obviously in the radical
of the center.

By construction, the $\idem_s$ and $\nilp^{\pm}_s$ have the properties
summarized in the following proposition.
\begin{prop}\label{prop-center}
  The center $\cZ$ of $\UresSL2$ at $\q=e^{\frac{i\pi}{p}}$ is
  $(3p\,{-}\,1)$-dimensional.  Its associative commutative algebra
  structure is described as follows: there are two ``special''
  primitive idempotents $\idem_0$ and $\idem_p$, \ $p\,{-}\,1$ other
  primitive idempotents $\idem_s$, $1\leq s\leq p-1$, and $2(p\,{-}\,1)$
  elements $\nilp^\pm_s$\ $(1\leq s\leq p\,{-}\,1)$ in the radical such
  that
  \begin{alignat*}{2}
    \idem_s\,\idem_{s'}&=\delta_{s,s'}\idem_s,
    &\quad &s,s'=0,\dots,p,\\
    \idem_s\,\nilp^\pm_{s'}&=\delta_{s,s'}\nilp^\pm_{s'},&
    \quad &0\leq s\leq p,~1\leq s'\leq p-1,\\
    \nilp^\pm_{s}\nilp^\pm_{s'}&=\nilp^\pm_{s}\nilp^\mp_{s'}
    =0,&
    \quad &1\leq s,s'\leq p-1.
  \end{alignat*}
\end{prop}
We call $\idem_s$, $\nilp^\pm_s$ the canonical basis elements in the
center, or simply the \textit{canonical central elements}.  They are
constructed somewhat more explicitly in~\bref{prop-center-explicit}.

We note that the choice of a bimodule isomorphism
$\mathsf{Reg}^*\to\mathsf{Reg}$ fixes the normalization of
the~$\nilp^\pm_s$.

\subsubsection{}\label{rem:coeffs} 
For any central element $A$ and its decomposition
\begin{equation}\label{decomp-general}
  A=\sum_{s=0}^{p} a_s \idem_s
  + \sum_{s=1}^{p-1} \bigl(c^+_s \nilp^+_s+c^-_s \nilp^-_s\bigr)
\end{equation}
with respect to the canonical central elements, \textit{the
  coefficient $a_s$ is the eigenvalue of $A$ in the irreducible
  representation $\repLambda(s)$}.  To determine the $c^+_s$ and
$c^-_s$ coefficients similarly, we fix the normalization of the basis
vectors as in~\bref{proj-mod-base}, i.e., such that $\nilp^+_s$ and
$\nilp^-_s$ act as
\begin{equation*}
  \nilp^+_s\,\toppr^{(+,s)}_n=\botpr^{(+,s)}_n,\quad
  \nilp^-_s\,\rightpr^{(-,s)}_k=\leftpr^{(-,s)}_k
\end{equation*}
in terms of the respective bases in the projective modules $\modL(s)$
and $\modP(p-s)$ defined in~\bref{module-L} and~\bref{module-P}.  Then
\textit{the coefficient $c^+_s$ is read off from the relation
  $A\toppr^{(+,s)}_n=c^+_s\botpr^{(+,s)}_n$ in $\modL(s)$, and
  $c^-_s$, similarly, from the relation
  $A\rightpr^{(-,s)}_k=c^-_s\leftpr^{(-,s)}_k$ in $\modP(p-s)$}.

\subsection{The Radford mapping for~$\UresSL2$}\label{sec:Radford-SL2}
For a Hopf algebra $A$ with a given cointegral, we recall the Radford
mapping $\radmap:A^*\to A$, see~\bref{sec:radford-all} (we use the hat
for notational consistency in what follows).  For $A=\UresSL2$, with
the cointegral $\coint$ in~\eqref{coint}, we are interested in the
restriction of the Radford mapping to the space of $q$-characters
$\Ch$ and, more specifically, to the image of the Grothendieck ring in
$\Ch$ via the mapping $\repX\mapsto\qTr_{\repX}$ (see~\eqref{qCh}).
We thus consider the mapping
\begin{align*}
  \Grring_{2p}\to{}&\cZ,\\
  \intertext{which acts on the irreducible representations as}
  \repX^{\pm}(s)\mapsto{}&
  \radmap{}^{\pm}(s)\equiv\radmap(\qTr_{\repX^{\pm}(s)})
  =\sum_{(\coint)} \Tr_{\repX^{\pm}(s)}(K^{p-1}\coint')\,\coint'',
  \quad 1\leq s\leq p.
\end{align*}
Let $\Radalg_{2p}$ be the linear span of the $\radmap{}^{\pm}(s)$
(the image of the Grothendieck ring in the center under the Radford
mapping).  As we see momentarily, $\Radalg_{2p}$ is $2p$-dimensional
and coincides with the algebra generated by the
$\radmap{}^{\,\alpha}(s)$.

It follows that
\begin{gather*}
  \radmap{}^{\,+}(1)=\coint,
\end{gather*}
in accordance with the fact that $\coint$ furnishes an embedding of
the trivial representation~$\repX^+(1)$ into~$\UresSL2$.  A general
argument based on the properties of the Radford mapping
(cf.~\cite{[L-center]}) and on the definition of the canonical
nilpotents $\nilp^{\pm}_s$ above implies that for
$s=1,\dots,p\,{-}\,1$, $\radmap{}^{\,+}(s)$ coincides with $\nilp^+_s$
up to a factor and $\radmap{}^{\,-}(s)$ coincides with $\nilp^-_{p-s}$
up to a factor.  We now give a purely computational proof of this
fact, which at the same time fixes the factors; we describe this in
some detail because similar calculations are used in what follows.

\begin{lemma}\label{lemma:phi-idem}
  For $1\leq s\leq p-1$,
  \begin{equation*}
    \radmap{}^{\,+}(s)=\omega_s \nilp^+_s,
    \quad
    \radmap{}^{\,-}(s)=\omega_{s} \nilp^-_{p-s},
    \qquad
    \omega_s = \ffrac{p\sqrt{2p}}{[s]^2}.
  \end{equation*}
  Also,
  \begin{equation*}
    \radmap{}^{\,+}(p) = p\sqrt{2p}\,\idem_{p},
    \qquad
    \radmap{}^{\,-}(p) = (-1)^{p + 1} p \sqrt{2p}\,\idem_{0}.
  \end{equation*}
  Therefore, the image of the Grothendieck ring under the Radford
  mapping is the socle (annihilator of the radical) of~$\cZ$.
\end{lemma}
\begin{proof}
  First, we recall~\eqref{coint} and use~\eqref{the-trace2}
  and~\eqref{Delta-formula} to evaluate
  \begin{equation}\label{phi-explicit}
    \radmap{}^{\,\alpha}( s) =
    \zeta \sum_{n=0}^{s - 1}
    \sum_{i=0}^{n}
    \sum_{j=0}^{2p - 1}
    \alpha^{ i + j} ([i]!)^2
    \q^{j(s - 1 - 2n)}\qbin{s - n + i - 1}{i}
    \qbin{n}{i} F^{p - 1 - i} E^{p - 1 - i} K^j
  \end{equation}
  (the calculation is very similar to the one in~\bref{prop-eval}).
  
  Next, we decompose $\radmap{}^{\,\alpha}(s)$ with respect to the
  canonical basis following the strategy in~\bref{rem:coeffs}.  That
  is, we use~\eqref{phi-explicit} to calculate the action of
  $\radmap{}^{\,+}(s)$ on the module $\modL(s')$ ($1\leq s'\leq p-1$).
  This action is nonzero only on the vectors $\toppr^{(+,s')}_n$
  (see~\bref{module-L}); because $\radmap{}^{\,+}(s)$ is central, it
  suffices to evaluate it on any single vector, which we choose
  as~$\toppr^{(+,s')}_0$.  For $1\leq s\leq p-1$,
  using~\eqref{eq:EmFm-prod2} and~\eqref{cas-action}, we then have
  \begin{multline}\label{radmap-acts}
    \radmap{}^{\,+}(s)\toppr^{(+,s')}_0
    = \zeta \sum_{n=0}^{s - 1}\!
    \sum_{i=0}^{n}\!
    \sum_{j=0}^{2p - 1}
    ([i]!)^2
    \q^{j(s + s' - 2 - 2n)}
    \qbin{s - n + i - 1}{i}
    \qbin{n}{i}\\*
    \shoveright{{}\times
      \prod_{r=0}^{p-2-i}
      \Bigl(\cas-\mfrac{\q^{2r+1}K+\q^{-2r-1}K^{-1}}{(\q-\q^{-1})^2}
      \Bigr)
      \toppr^{(+,s')}_0}\\
    {}= \zeta \sum_{n=0}^{s - 1}\!
    \sum_{i=0}^{n}\!
    \sum_{j=0}^{2p - 1}
    (-1)^{p+i}
    ([i]!)^2
    \q^{j(s + s' - 2 - 2n)}
    \qbin{s - n + i - 1}{i}
    \qbin{n}{i}
    \prod_{r=1}^{p-2-i}\!
    [s'+r][r]\,
    \botpr^{(+,s')}_0,
  \end{multline}
  with the convention that whenever $p-2-i=0$, the product over~$r$
  evaluates as~$1$.  We simultaneously see that the diagonal part of
  the action of~$\radmap{}^{\,+}(s)$ on~$\modL(s')$ vanishes.
  
  Analyzing the cases where the product over~$r$
  in~\eqref{radmap-acts} involves $[p]=0$, it is immediate to see that
  a necessary condition for the right-hand side to be nonzero is
  $s'\leq s$. Let therefore $s=s'+\ell$, where $\ell\geq0$.  It is
  then readily seen that~\eqref{radmap-acts} vanishes for odd~$\ell$;
  we thus set $\ell=2m$, which allows us to evaluate 
  \begin{multline*}
    \radmap{}^{\,+}(s'+2m)\,\toppr^{(+,s')}_0={}\\
    {}=2p \zeta\sum_{i=s'-1}^{m+s'-1}(-1)^{p+i}([i]!)^2
    \qbin{m+i}{i}\qbin{m+s'-1}{i}\ffrac{[p-2-i+s']!}{[s']!}\,
    [p-2-i]!\,\botpr^{(+,s')}_0.
  \end{multline*}
  But this vanishes for all $m>0$ in view of the identity
  \begin{gather*}
    \sum_{j=0}^m(-1)^j\ffrac{[j+s'+1]\dots[j+s'+m-1]}{[j]![m-j]!}
    =\ffrac{1}{[m]}\sum_{j\in\oZ}(-1)^j\qbin{m}{j}\qbin{m+s'-1+j}{m-1}
    =0,
    \quad
    m\geq1.
  \end{gather*}
  
  Thus, $\radmap{}^{\,+}(s)$ acts by zero on $\modL(s')$ for all $s'\neq
  s$; it follows similarly that $\radmap{}^{\,+}(s)$ acts by zero on
  $\modP(s')$ for all~$s'$ and on both Steinberg modules
  $\repX^{\pm}(p)$.  Therefore, $\radmap{}^{\,+}(s)$ is necessarily
  proportional to~$\nilp^+_s$, with the proportionality coefficient to
  be found from the action on $\modL(s)$.  But for $s'=s$, the sum
  over $j$ in the right-hand side of~\eqref{radmap-acts} is zero
  unless $n=s-1$, and we have
  \begin{align*}
    \radmap{}^{\,+}(s)\toppr^{(+,s)}_0
    &= \ffrac{2p\,\zeta}{[s]}
    \sum_{i=0}^{s-1}
    (-1)^{p+i}
    [i]!\ffrac{[p-2-i]![s+p-2-i]!}{[s-1-i]!}\,    
    \botpr^{(+,s)}_0,
    \intertext{where the terms in the sum are readily seen to vanish
      unless $i=s-1$, and therefore}
    &=2p\,\zeta\,(-1)^{p+s+1}\ffrac{[p-1]!\,[s-1]!\,[p-1-s]!}{[s]}\,
    \botpr^{(+,s)}_0,
    \end{align*}
    which gives $\omega_s$ as claimed.  The results for
    $\radmap{}^{\,-}(s)$ ($1\leq s\leq p-1$) and $\radmap{}^{\,\pm}(p)$
    are established similarly.
\end{proof}

\medskip

It follows (from the expression in terms of the canonical central
elements; cf.~\cite{[L-center]} for the small quantum group) that the
two images of the Grothendieck ring in the center, $\Drinalg_{2p}$ and
$\Radalg_{2p}$, span the entire center:
 \begin{equation*}
   \Drinalg_{2p}\cup\Radalg_{2p}=\cZ.
 \end{equation*}
 We next describe the intersection of the two Grothendieck ring images
 in the center (cf.~\cite{[L-center]} for the small quantum group).
 This turns out to be the Verma-module ideal
 (see~\bref{Verma-in-center}).

\begin{prop}\label{prop:phi+phi}  
  $\Drinalg_{2p}\cap\Radalg_{2p}=\Videal_{p+1}$.  
\end{prop}
\begin{proof}
  Proceeding similarly to the proof of~\bref{lemma:phi-idem}, we
  establish the formulas
  \begin{multline}\label{phi+phi}
    \radmap{}^{\,+}(s) + \radmap{}^{\,-}( p - s)
    = \zeta\ffrac{([p - 1]!)^2\!}{p}\\*
    {}\times\Bigl((-1)^{p-s}\vvarkappa(0) + \sum_{s'=1}^{p-1}
    (-1)^{p+s+s'}\bigl(\q^{s s'} + \q^{-s s'}\bigr) \vvarkappa(s')
    + \vvarkappa(p)\Bigr)    
  \end{multline}
  for $s=1,\dots,p-1$, and
  \begin{equation}\label{phi-p}
    \begin{split}
      \radmap{}^{\,+}(p) &= \ffrac{1}{\sqrt{2p}}\Bigl(\vvarkappa(0) +
      2\sum_{s'=1}^{p-1}\vvarkappa(p-s') + \vvarkappa(p)\Bigr),\\
      \radmap{}^{\,-}(p) &=
      \ffrac{1}{\sqrt{2p}}\Bigl((-1)^{p}\vvarkappa(0) + 
      2\sum_{s'=1}^{p-1}(-1)^{s'}\vvarkappa(p-s') + \vvarkappa(p)\Bigr),
    \end{split}
  \end{equation}
  which imply the proposition.  The derivation may in fact be
  simplified by noting that as a consequence of~\eqref{pi+pi} and
  \bref{remarks-D}(\ref{item:w}), $\radmap{}^{\,+}(s) + \radmap{}^{\,-}( p
  - s)$ belongs to the subalgebra generated by the Casimir element,
  which allows using~\eqref{P(C)-idem}.  
\end{proof}

\subsection{The $\UresSL2$ ribbon element}\label{SL2-ribbon} We
finally recall (see~\bref{sec:ribbon} and~\cite{[RSts]}) that a ribbon
element $\ribbon\,{\in}\,A$ in a Hopf algebra $A$ is an invertible
central element satisfying~\eqref{def-ribbon}.  For $\UresSL2$, the
ribbon element is actually given in~\eqref{rib-bar-D}, rewritten~as
\begin{equation*}
  \ribbon =\ffrac{1-i}{2\sqrt{p}}\sum_{m=0}^{p-1}\sum_{j=0}^{2p-1}
  \ffrac{(\q-\q^{-1})^m}{[m]!}\,
  \q^{-\frac{m}{2}+mj+\half(j+p+1)^2} F^m E^m K^{j}
\end{equation*}
in terms of the $\UresSL2$ generators.  A calculation similar to the
one in the proof of~\bref{lemma:phi-idem} shows the following
proposition.
\begin{prop}\label{ribbon-basis}
  The $\UresSL2$ ribbon element is decomposed in terms of the
  canonical central elements as
  \begin{align*}
    \ribbon
    &=\sum_{s=0}^{p}(-1)^{s+1} \q^{-\half(s^2-1)}\idem_s
    +\sum_{s=1}^{p - 1} (-1)^p \q^{-\half(s^2 - 1)}
    [s]\,\ffrac{\q - \q^{-1}}{\sqrt{2 p}}\,\vvarphi(s),
  \end{align*}
  where 
  \begin{gather}\label{hat-varphi}
    \vvarphi(s) =
    \ffrac{p-s}{p}\,\radmap{}^{\,+}(s) - \ffrac{s}{p}\,\radmap{}^{\,-}(p-s),
    \quad 1\leq s\leq p-1.
  \end{gather}
\end{prop}
Strictly speaking, expressing $\ribbon$ through the canonical central
elements requires using~\bref{lemma:phi-idem}, but below we need
$\ribbon$ expressed just through $\radmap{}^{\pm}(s)$.

\section{$\SLiiZ$-representations on the center
  of~$\UresSL2$}\label{sec:SLiiZ-restr} In this section, we first
recall the standard $\SLiiZ$-action~\cite{[Lyu],[LM],[Kerler]}
reformulated for the center $\cZ$ of~$\UresSL2$.  Its definition
involves the ribbon element and the Drinfeld and Radford mappings.
{}From the multiplicative Jordan decomposition for the ribbon element,
we derive a factorization of the standard
$\SLiiZ$-representation~$\repLy$, \ 
$\repLy(\gamma)=\repSw(\gamma)\repA(\gamma)$, where $\repSw$ and
$\repA$ are also $\SLiiZ$-representations on~$\cZ$.  We then establish
the equivalence to the $\SLiiZ$-representation on $\cZ_{\mathrm{cft}}$
in~\bref{mod-on-char}.

\subsection{The standard $\SLiiZ$-representation on~$\cZ$}
Let $\repLy$ denote the $\SLiiZ$-represen\-tation on the center $\cZ$
of~$\UresSL2$ constructed, as a slight modification of the
representation in~\cite{[Lyu],[LM],[Kerler]}, as follows.  We let
$\modS\equiv\repLy(S):\cZ\to\cZ$ and $\modT\equiv\repLy(T):\cZ\to\cZ$
be defined as
\begin{equation}\label{TS-def}
  \modS(a) =
  \radmap\bigl(\drmap^{-1}(a)\bigr),
  \quad
  \modT(a)=b\,\modS^{-1}\bigl(\ribbon^{-1}\bigl(\modS(a)\bigr)\bigr),
  \qquad a\in\cZ,
\end{equation}
where $\ribbon$ is the ribbon element, $\drmap$ is the Drinfeld
mapping, $\radmap$ is the Radford mapping, and $b$ is the
normalization factor
\begin{equation*}
  b=e^{i\pi(\frac{(p+1)^2}{2p}-\frac{1}{12})}.
\end{equation*}
We call it the \textit{standard $\SLiiZ$-representation}, to
distinguish it from other representations introduced in what follows.

We recall that $\modS^2$ acts via the antipode on the center of the
quantum group, and hence acts identically on the center of~$\UresSL2$,
\begin{equation}\label{S2}
  \modS^2=\id_{\cZ}.
\end{equation}

\begin{Thm}\label{Thm:equiv}
  \addcontentsline{toc}{subsection}{\thesubsection. \ \ Equivalence
    theorem} The standard $\SLiiZ$-representation on the center $\cZ$
  of~$\,\UresSL2$ at $\q=e^{i\pi/p}$ is equivalent to the
  $(3p\,{-}\,1)$-dimensional $\SLiiZ$-representation on
  $\cZ_{\mathrm{cft}}$ \textup{(}the extended characters of the
  $(1,p)$ conformal field theory model
  in~\bref{mod-on-char}\textup{)}.
\end{Thm}
We therefore abuse the notation by letting $\repLy$ denote both
representations.
\begin{proof}
  We introduce a basis in $\cZ$ as
  \begin{align*}
    &\rrho(s),\quad 1\leq s\leq p-1,\\
    &\vvarkappa(s),\quad 0\leq s\leq p,\\
    &\PPhi(s),\quad 1\leq s\leq p-1,
  \end{align*}
  where
  \begin{gather*}
    \rrho(s) =
    \ffrac{p-s}{p}\,\cchi^{+}(s) - \ffrac{s}{p}\,\cchi^{-}(p-s),
  \end{gather*}
  $\vvarkappa(s)$ are defined in~\eqref{verma-def}, and
  \begin{equation*}
    \PPhi(s)
    =\ffrac{1}{\sqrt{2p}}
    \sum_{r=1}^{p-1}(-1)^{r+s+p}(\q^{r s} - \q^{-r s})
    \vvarphi(r)
  \end{equation*}
  (with $\vvarphi(s)$ defined in~\eqref{hat-varphi}).  That this is a
  basis in the center follows, e.g., from the decomposition into the
  canonical central elements.
  
  The mapping
  \begin{align*}
    \rho_{s\andp}&\mapsto\rrho(s),\quad 1\leq s\leq p-1,\\
    \varkappa_{s\andp}&\mapsto\vvarkappa(s),\quad 0\leq s\leq p,\\
    \varphi_{s\andp}&\mapsto\PPhi(s),\quad 1\leq s\leq p-1
  \end{align*}
  between the bases in $ \cZ_{\mathrm{cft}}$ and in~$\cZ$ establishes
  the equivalence.  Showing this amounts to the following checks.
  
  First, we evaluate $\modS (\rrho(s))$ as
  \begin{multline*}
    \modS (\rrho(s)) =
    \radmap\circ\drmap^{-1}(\ffrac{p-s}{p}\,\cchi^{+}(s)
    - \ffrac{s}{p}\,\cchi^{-}(p-s))\\*
    {}=
    \ffrac{p-s}{p}\,\radmap{}^{\,+}(s) - \ffrac{s}{p}\,\radmap{}^{\,-}(p-s)
    = \vvarphi(s),
  \end{multline*}
  and hence, in view of~\eqref{S2},
  \begin{equation}\label{S-rho-phi-1}
    \modS (\vvarphi(s)) = \rrho(s),\quad 1\leq s\leq p-1.
  \end{equation}
  We also need this formula rewritten in terms of
  \begin{equation*}
    \hrho(r)= \ffrac{1}{\sqrt{2p}}
    \sum_{s=1}^{p-1}(-1)^{r + s + p}(\q^{rs}-\q^{-rs})\rrho(s),
  \end{equation*}
  that is,
  \begin{equation}\label{S-rho-phi-2}
    \modS (\PPhi(s)) = \hrho(s),\quad 1\leq s\leq p-1.
  \end{equation}
  
  Further, we use \eqref{phi+phi} and \eqref{phi-p} to evaluate
  $\modS(\vvarkappa(s))$ as
  \begin{multline*}
    \modS (\vvarkappa(s)) =
    \radmap\circ\drmap^{-1}(\cchi^{+}(s) + \cchi^{-}(p-s))
    = \radmap{}^{\,+}(s) + \radmap{}^{\,-}(p-s) =\\
    = \ffrac{1}{\sqrt{2p}}\Bigl((-1)^{p-s}\vvarkappa(0)
    + \sum_{s'=1}^{p-1} (-1)^{s'}\bigl(\q^{s s'}
    + \q^{-s s'}\bigr) \vvarkappa(p-s')
    + \vvarkappa(p)\Bigr),\quad 0\leq s\leq p,
  \end{multline*}
  where we set $\cchi^{\pm}(0)=\radmap{}^{\,\pm}(0)=0$.  This shows that
  $\modS$ acts on $\rrho(s)$, $\vvarkappa(s)$, and $\PPhi(s)$ as on
  the respective basis elements $\rho_s$, $\varkappa_s$, and
  $\varphi_s$ in~$\cZ_{\mathrm{cft}}$.

  Next, it follows from~\bref{ribbon-basis} that~$\ribbon$ acts on
  $\radmap{}^{\,\pm}(s)$ as
  \begin{equation*}
    \begin{aligned}
      \ribbon\radmap{}^{\,+}(s)
      &=
      (-1)^{s+1} \q^{-\half(s^2-1)} \radmap{}^{\,+}(s),
      \\
      \ribbon\radmap{}^{\,-}(s)
      &=
      (-1)^{p+1}\q^{-\half(p^2 + s^2-1)} \radmap{}^{\,-}(s),
    \end{aligned}
    \quad 1\leq s\leq p.
  \end{equation*}
  As an immediate consequence, in view of
  $\modT\drmap^{\pm}(s)=b\modS^{-1}(\ribbon^{-1}\radmap{}^{\,\pm}(s))$,
  we have
  \begin{equation}\label{modT-acts}
    \modT\drmap^{+}(s) =
    \lambda_{p,s}\drmap^{+}(s),
    \quad
    \modT\drmap^{-}(s) =
    \lambda_{p,p-s}\drmap^{-}(s),
    \qquad 1\leq s\leq p,
  \end{equation}
  where $\lambda_{p,s}$ is defined in~\eqref{eq:lambda}.  It follows
  that $\modT$ acts on $\rrho(s)$ and $\vvarkappa(s)$ as on the
  respective basis elements $\rho_s$ and $\varkappa_s$
  in~$\cZ_{\mathrm{cft}}$.
  
  Finally, we evaluate $\modT\PPhi(s)$.  Recalling \bref{ribbon-basis}
  to rewrite~$\ribbon$ as
  \begin{equation*}
    \ribbon=\sum_{t=0}^{p}(-1)^{t+1}\q^{-\half(t^2-1)}
    \idem_t(\one+\PPhi(1)),
  \end{equation*}
  we use~\eqref{S2} and~\eqref{S-rho-phi-2}, with the result
  \begin{equation*}
    \modT \PPhi(s)
    = b\modS\,\ribbon^{-1}\,
    \hrho(s)
    = b\modS\, \sum_{t=0}^{p}(-1)^{t+1} \q^{\half(t^2-1)}\idem_t
    \,\bigl(\one - \PPhi(1)\bigr)
    \hrho(s).
  \end{equation*}
  But (a simple rewriting of the formulas in~\bref{eigenP})
  \begin{equation*}
    \hrho(s)   = (-1)^{p+s}\ffrac{\sqrt{2p}}{\q^{s} -
    \q^{-s}}\Bigl(\idem_s - \ffrac{\q^{s} +
    \q^{-s}}{[s]^2}\,\nilp_s\Bigr),
  \end{equation*}
  and therefore (also recalling the projector properties to see that
  only one term survives in the sum over~$t$)
  \begin{multline*}
    \modT\PPhi(s) = -b\ffrac{\sqrt{2p}}{\q^{s} - \q^{-s}}\,\modS\,
    \sum_{t=0}^{p}(-1)^{t+s+p} \q^{\half(t^2-1)}\idem_t
    \,\bigl(\one - \PPhi(1)\bigr)
    \Bigl(\idem_s -  \ffrac{\q^{s} + \q^{-s}}{[s]^2}\,\nilp_s\Bigr)={}\\
    = -b\ffrac{\sqrt{2p}}{\q^{s} - \q^{-s}}\,\modS\, (-1)^{p}
    \q^{\half(s^2-1)}\idem_s \Bigl(\idem_s - \ffrac{\q^{s} +
      \q^{-s}}{[s]^2}\,\nilp_s
    - \PPhi(1)\idem_s\Bigr)\\
    = b(-1)^{s+1}\q^{\half(s^2-1)}\, \modS \,\hrho(r)
    +b\ffrac{(-1)^{p}\sqrt{2p}\,\q^{\half(s^2-1)}}{\q^{s} - \q^{-s}}\,
    \modS\, \PPhi(1)\idem_s.
  \end{multline*}
  Here, $\modS\hrho(r)=\PPhi(r)$ and $\PPhi(1)\idem_s
  =(-1)^{s+p+1}\,\ffrac{\q^s-\q^{-s}}{\sqrt{2p}}\,\vvarphi(s)$, and
  hence
  \begin{equation*}
    \modT \PPhi(s)  
    = \lambda_{p,s}\bigl( \PPhi(s) + \rrho(s)\bigr).
  \end{equation*}  
  This completes the proof.
\end{proof}

\subsection{Factorization of the standard $\SLiiZ$-representation on
  the center}\label{two-rep-on-Z} In view of the equivalence of
representations, the $\SLiiZ$-representation $\repLy$ on the center
admits the factorization established in~\bref{thm:R-decomp}.
Remarkably, this factorization can be described in ``intrinsic''
quantum-group terms, as we now show.  That is, we construct two more
$\SLiiZ$-representations on~$\cZ$ with the properties described
in~\bref{thm:modular-2}.

\subsubsection{} For the ribbon element $\ribbon$, we consider its
multiplicative Jordan decomposition
\begin{equation}\label{ribbon-factor}
  \ribbon=\ribbon^*\bar\ribbon
\end{equation}
into the semisimple part
\begin{equation*}
  \bar\ribbon=\sum_{s=0}^{p}(-1)^{s+1} \q^{-\half(s^2-1)}\idem_s
\end{equation*}
and the unipotent part
\begin{equation*}
  \ribbon^*= \one + \PPhi(1).
\end{equation*}
With \eqref{ribbon-factor}, we now let $\modT^*:\cZ\to\cZ$ and
$\bar\modT:\cZ\to\cZ$ be defined by the corresponding parts of the
ribbon element, similarly to~\eqref{TS-def}:
\begin{equation*}
  \modT^*(a) = \modS^{-1}\bigl({\ribbon^*}^{-1}\modS(a)\bigr),
  \quad
  \bar\modT(a) = b\modS^{-1}\bigl({\bar\ribbon}^{-1}\modS(a)\bigr),
  \qquad
  a\in\cZ.
\end{equation*}
Then, evidently,
\begin{equation*}
  \modT=\modT^*\bar\modT.
\end{equation*}

\subsubsection{}
We next define a mapping $\amap:\UresSL2^*\to\UresSL2$ as
\begin{equation}\label{amap-def}
  \amap(\beta)=(\beta\tensor\id)(\cointa),
\end{equation}
where
\begin{equation*}
  \cointa=(\ribbon^*\tensor\ribbon^*)\Delta(\modS(\ribbon^*)).
\end{equation*}
It intertwines the coadjoint and adjoint actions of $\UresSL2$, and we
therefore have the mapping $\amap:\Ch(\UresSL2)\to\cZ$, which is
moreover an isomorphism of vector spaces.  We~set
\begin{equation}\label{S-pieces}
  \modS^*=\radmap\circ\amap^{-1},
  \qquad
  \bar\modS=\amap\circ\drmap^{-1}.
\end{equation}
This gives the decomposition
\begin{equation*}
  \modS=\modS^*\bar\modS.
\end{equation*}

\begin{thm}\label{thm:factorization}
  The action of $\;\modS^*$ and $\;\modT^*$ on the center generates
  the $\SLiiZ$-represen\-ta\-tion $\repA$, and the action of
  $\bar\modS$ and $\bar\modT$ on the center generates the
  $\SLiiZ$-representation $\repSw$, such that
  \begin{enumerate}
  \item $\repSw(\gamma)\repA(\gamma') =\repA(\gamma')\repSw(\gamma)$
    for all $\gamma,\gamma'\in\SLiiZ$,
    
  \item the representation $\repSw$ restricts to the Grothendieck ring
    (i.e., to its isomorphic image in the center), and
    
  \item $\repLy(\gamma)=\repSw(\gamma)\repA(\gamma)$ for all
    $\gamma\in\SLiiZ$,
\end{enumerate}
and $\repLy$ and $\repSw$ are isomorphic to the respective
$\SLiiZ$-representations on $\cZ_{\mathrm{cft}}$
in~\bref{thm:R-decomp}.
\end{thm}
The verification is similar to the proof of~\bref{Thm:equiv}, with
\begin{equation*}
  {\modS^*}^{-1}\bigl(\radmap{}^{\pm}(s)\bigr)
  =\amap(\qTr_{\repX^{\pm}(s)})
  =(\Tr_{\repX^{\pm}(s)}\tensor\id)\bigl((K^{p-1}\tensor\one)N\bigr)
\end{equation*}
and
\begin{equation*}
  {\modS^*}^{-1}\bigl(\drmap^{\pm}(s)\bigr)
  =(\rint\tensor\id)\bigl(S(\drmap^{\pm}(s))\tensor\one)N\bigr)
\end{equation*}
(and similarly for $\bar\modS$), based on the
formula\enlargethispage{\baselineskip}
\begin{equation*}
  \modS(\ribbon^*)=\modS(\one + \PPhi(1))
  = \radmap{}^{\,+}(1) + \hrho(1)
  =\coint + \hrho(1).
\end{equation*}

\subsubsection{} The three mappings involved
in~\eqref{S-pieces}\,---\,$\radmap$ defined in~\eqref{radford-def},
$\drmap$ defined in~\eqref{drinfeld-def}, and~$\amap$
in~\eqref{amap-def}\,---\,can be described in a unified way as
follows.  Let $A$ be a ribbon Hopf algebra endowed with the standard
$\SLiiZ$-representation.  For $x\,{\in}\, A$, we define
\begin{equation*}
  \llambda_x:A^*\to A\pagebreak[3]
\end{equation*}\pagebreak[3]
as
\begin{equation*}
  \llambda_x(\beta)=(\beta\tensor\id)\bigl(
  (x\tensor x)\Delta(\modS(x))\bigr),
\end{equation*}
where $\modS$ is the standard action of~$\left(\begin{smallmatrix}
    0&1\\
    -1&0
  \end{smallmatrix}\right)$.  Taking $x$ to be the three
elements $\one$, $\ribbon$, and $\ribbon^*$, we have
\begin{equation*}
  \llambda_{\one}=\radmap,\qquad
  \llambda_{\ribbon}=\drmap,\qquad
  \llambda_{\ribbon^*}=\amap.
\end{equation*}


\section{Conclusions}

We have shown that the Kazhdan--Lusztig correspondence, understood in
a broad sense as a correspondence between conformal field theories and
quantum groups, extends into the nonsemisimple realm such that a
number of structures on the conformal field theory side and on the
quantum group side are actually isomorphic, which signifies an
``improvement'' over the case of rational$/$semisimple conformal field
theories.

Although much of the argument in this paper is somewhat too
``calculational,'' and hence apparently ``accidental,'' we hope that a
more systematic derivation can be given.  In fact, the task to place
the structures encountered in the study of nonsemisimple Verlinde
algebras into the categorical context~\cite{BK,fuRs4,fuRs8,KElu} was
already formulated in~\cite{[FHST]}.  With the quantum-group
counterpart of nonsemisimple Verlinde algebras and of the
$\SLiiZ$-representations on the conformal blocks studied in this paper
in the $(1,p)$ example, this task becomes even more compelling.

We plan to address Claim~\ref{item:equiv-cat} of the Kazhdan--Lusztig
correspondence (see page~\pageref{item:equiv-cat}) between the
representation categories of the $\algW(p)$ algebra and
of~$\UresSL2$~\cite{[FGST2]}.  This requires constructing
vertex-operator analogues of extensions among the irreducible
representations (generalizing the $(1,2)$ case studied
in~\cite{[FFHST]}).

Another direction where development is welcome is to go over from
$(1,p)$ to $(p',p)$ models of logarithmic conformal field theories,
starting with the simplest such model, $(2,3)$, whose content as a
minimal theory is trivial, but whose logarithmic version may be quite
interesting.

\subsubsection*{Acknowledgments} 
We are grateful to A.~Belavin, E.~Feigin, M.~Fin\-kel\-berg, K.~Hori,
B.~Khesin, S.~Loktev, S.~Parkhomenko, Y.~Soibelman, M.A.~Soloviev, and
B.L.~Voronov for useful discussions.  This paper was supported in part
by the RFBR Grants 04-01-00303 (BLF, AMG, AMS, and IYT),
LSS-1578.2003.2 (AMS and IYT), 02-01-01015 and LSS-2044.2003.2 (BLF),
INTAS Grant 03-51-3350 (BLF).  AMS is grateful to the Fields
Institute, where a part of this paper was written, for hospitality.

\appendix

\section{Hopf algebra definitions and standard facts} \label{app:Hopf}
We let $A$ denote a Hopf algebra with comultiplication~$\Delta$,
counit~$\epsilon$, and antipode~$S$.  The general facts summarized
here can be found
in~\cite{[LSw],[Rad-antipode],[Drinfeld],[Kassel],[ChP]}.

\subsection{Adjoint and coadjoint actions, center, and
  $q$-characters}\label{sec:q-chars} For a Hopf algebra $A$, the
adjoint and coadjoint actions $\ad_a:A\to A$ and $\ad^*_a:A^*\to A^*$
($a\,{\in}\, A$) are defined~as
\begin{equation*}
  \ad_a(x)=\sum_{(a)} a'xS(a''),\quad
  \ad^*_a(\beta)=\beta\bigl(\sum_{(a)} S(a')?a''\bigr),\quad
  a,x\in A,\quad\beta\in A^*.
\end{equation*}

The center $\cZ(A)$ of $A$ can be characterized as the set
\begin{equation*}
  \cZ(A)=\bigl\{ y\in A \bigm| \ad_x(y)
  =\epsilon(x)y\quad \forall x\in A\bigr\}.
\end{equation*}

By definition, the space $\Ch(A)$ of $q$-characters is
\begin{multline}\label{Ch-def}
  \Ch(A)=\bigl\{\beta\in A^* \bigm| \ad^*_x(\beta)
  =\epsilon(x)\beta\quad \forall x\in A\bigr\}\\*
  = \bigl\{\beta\in A^* \bigm| \beta(xy)=\beta\bigl(S^2(y)x\bigr)
  \quad \forall x,y\in A\bigr\}.
\end{multline}

Given an invertible element $t\,{\in}\, A$ satisfying
$S^2(x)=txt^{-1}$ for all $x\,{\in}\, A$, we define the linear mapping
$\qtr^t_V:A\to\oC$ for any $A$-module $\repX$ as
\begin{equation}\label{q-trace}
  \qtr^t_{\repX} =\tr_{\repX}(t^{-1}?).
\end{equation}
\begin{lemma}[\cite{[ChP],[Kassel]}]\label{lemma:qch}
  For any $A$-module $\repX$ and an element $t$ such that
  $S^2(x)=txt^{-1}$, we have
  \begin{enumerate}
  \item $\qtr^t_{\repX}\in \Ch(A)$
  \item if in addition $t$ is group-like, i.e., $\Delta(t)=t\tensor
    t$, then 
    \begin{equation*}
      \qtr^t:\repX\mapsto\qtr^t_{\repX}(?)
    \end{equation*}
    is a homomorphism of the Grothendieck ring to the ring of
    $q$-characters.
  \end{enumerate}
\end{lemma}

\subsection{(Co)integrals, comoduli, and balancing}\label{app:int}
For a Hopf algebra $A$, a \textit{right integral}~$\rint$ is a linear
functional on $A$ satisfying
\begin{equation*}
  (\rint\tensor\id)\Delta(x)=\rint(x)\one
\end{equation*}
for all $x\,{\in}\, A$.  Whenever such a functional exists, it is
unique up to multiplication with a nonzero constant.

A \textit{comodulus}~$\comodul$ is an element in $A$ such that
\begin{equation*}
  (\id\tensor\rint)\Delta(x)
  =\rint(x)\comodul.
\end{equation*}
The left--right \textit{cointegral}~$\coint$ is an element in $A$ such
that
\begin{equation*}
  x\coint=\coint x =\epsilon(x)\coint,\quad\forall x\in A.
\end{equation*}
If it exists, this element is unique up to multiplication with a
nonzero constant.  We also note that the cointegral gives an embedding
of the trivial representation of $A$ in the bimodule~$A$.  We use the
normalization $\rint(\coint)=1$.

Whenever a square root of the comodulus $\comodul$ can be calculated
in a Hopf algebra $A$, the algebra admits the \textit{balancing
  element} $\balance$ that satisfies
\begin{equation}\label{balance-prop}
  S^2(x)=\balance x\balance^{-1},\quad
  \Delta(\balance)=\balance\tensor\balance,
\end{equation}
In fact, we have the following lemma.
\begin{lemma}[\cite{[Drinfeld]}]
  \begin{equation}\label{bal-comod}
  \balance^2=\comodul.
  \end{equation}
\end{lemma}

\subsection{The Radford mapping}\label{sec:radford-all} Let $A$ be a
Hopf algebra with the right integral $\rint$ and the left--right
cointegral~$\coint$.  The Radford mapping $\radmap:A^*\to A$ and its
inverse $\radmap{}^{-1}:A\to A^*$ are given by
\begin{equation}\label{radford-def}
  \radmap(\beta)
  =\sum_{(\coint)}\beta(\coint')\coint'',
  \quad
  \radmap{}^{-1}(x)=\rint(S(x)?).
\end{equation}
\begin{lemma}[\cite{[Swe],[Rad]}]\label{lemma:rad-map}
  $\radmap$ and $\radmap{}^{-1}$ are inverse to each other,
  $\radmap\radmap{}^{-1}=\id_{A}$, $\radmap{}^{-1}\radmap=\id_{A^*}$, and
  intertwine the left actions of $A$ on $A$ and $A^*$, and similarly
  for the right actions.
\end{lemma}
Here, the left-$A$-module structure on $A^*$ is given by
$a\acts\beta=\beta(S(a)?)$ (and on $A$, by the regular action).

\subsection{Quasitriangular Hopf algebras and the $R$ and $M$
  matrices}\label{app:quasitriangle}
\subsubsection{$R$-matrix}
A quasitriangular Hopf algebra $A$ has an invertible element
$R\,{\in}\, A\tensor A$ satisfying
\begin{gather}
  \Delta^{\mathrm{op}}(x)=R\Delta(x) R^{-1},\label{eq:def-prop-R-d}
  \\
  (\Delta\otimes\id)(R)= R_{13} R_{23},\label{eq:quasi-1}
  \\
  (\id\otimes\Delta)( R)= R_{13} R_{12},\label{eq:quasi-2-d}
  \\
  R_{12}R_{13}R_{23}=R_{23}R_{13}R_{12},\notag\\
  (\epsilon\tensor\id)(R)=\one=(\id\tensor\epsilon)(R),\notag\\
  (S\tensor S)(R)=R.\notag
\end{gather}

\subsubsection{$M$-matrix}\label{app:M}
For a quasitriangular Hopf algebra $A$, the $M$-matrix is defined as
\begin{equation*}
  M=R_{21}R_{12}\in A\tensor A.
\end{equation*}
It satisfies the relations
\begin{align}\label{M-Delta}
  (\Delta\tensor\id)(M)&=R_{32}M_{13}R_{23} ,\\
  M\Delta(x)&=\Delta(x) M\quad \forall x\in A.
  \label{M-commutes}  
\end{align}
Indeed, using \eqref{eq:quasi-2-d}, we find
$(\Delta\otimes\id)(R_{21})= R_{32} R_{31}$ and then
using~\eqref{eq:quasi-1}, we obtain~\eqref{M-Delta}.  Next,
from~\eqref{eq:def-prop-R-d}, which we write as
$R_{12}\Delta(x)=\Delta^{\mathrm{op}}(x)R_{12}$, it follows that
$R_{21}R_{12}\Delta(x) =(R_{12}\Delta(x))^{\mathrm{op}}R_{12}
=(\Delta^{\mathrm{op}}(x)R_{12})^{\mathrm{op}}R_{12}
=\Delta(x)R_{21}R_{12}$, that
is,~\eqref{M-commutes}.\enlargethispage{\baselineskip}

If in addition~$M$ can be represented as
\begin{equation}\label{M-factorizable}
  M=\sum_I \pbwd_I\tensor \pbwdd_I,
\end{equation}
where~$\pbwd_I$ and~$\pbwdd_I$ are two \textit{bases} in~$A$, the Hopf
algebra~$A$ is called \textit{factorizable}.

\subsubsection{The square of the antipode~\cite{[Drinfeld],[Lyu]}}
In any quasitriangular Hopf algebra, the square of the antipode is
represented by a similarity transformation
\begin{equation*}
  S^2(x)=  \sqs x\sqs^{-1}
\end{equation*}
where the \textit{canonical element} $\sqs$ is given by
\begin{gather}\label{canon-sqs}
  \sqs= \cdot\bigl((S\tensor\id)R_{21}\bigr),\quad
  \sqs^{-1}=\cdot\bigl((S^{-1}\tensor S)R_{21}\bigr)
\end{gather}
(where $\cdot(a\tensor b)=ab$) and satisfies the property
\begin{gather}\label{Delta-u}
  \Delta(\sqs)= M^{-1}(\sqs\tensor\sqs)=(\sqs\tensor\sqs)M^{-1}.
\end{gather}

Any invertible element $t$ such that $S^2(x)=txt^{-1}$ for all $x\in
A$ can be expressed as $t=\theta\sqs$, where $\theta$ is an invertible
central element.

\subsection{The Drinfeld mapping}\label{sec:Drpdef}Given an
$M$-matrix (see~\bref{app:M}), we define the Drinfeld mapping
$\drmap:A^*\to A$ as
\begin{equation}\label{drinfeld-def}
  \drmap(\beta)=(\beta\tensor\id)M
  =\sum_I\beta(\pbwd_I)\pbwdd_I.
\end{equation}
\begin{lemma}[\cite{[Drinfeld]}]\label{lemma:Dr-map}
  In a factorizable Hopf algebra $A$, the Drinfeld mapping
  $\drmap:A^*\to A$ intertwines the adjoint and coadjoint actions of
  $A$ and its restriction to the space $\Ch$ of $q$-characters gives
  an isomorphism of associative algebras
  \begin{equation*}
    \Ch(A)\xrightarrow{\sim}\cZ(A).
  \end{equation*}
\end{lemma}

\subsection{Ribbon algebras}\label{sec:ribbon} A \textit{ribbon
  Hopf algebra}~\cite{[RSts]} is a quasitriangular Hopf algebra
equipped with an invertible central element $\ribbon$, called the
\textit{ribbon element}, such that
\begin{equation}\label{def-ribbon}
   \ribbon^2= \sqs S(\sqs),\quad
 S(\ribbon)=\ribbon,\quad\epsilon(\ribbon)=1,
  \quad
  \Delta(\ribbon)=M^{-1}(\ribbon\tensor\ribbon).
\end{equation}

In a ribbon Hopf algebra, 
\begin{equation}\label{balance-ribbon}
  \balance=\ribbon^{-1}\sqs,
\end{equation}
where $\balance$ is the balancing element (see~\bref{app:int}).

\subsubsection{}\label{app:qCh}Let $A$ be a ribbon Hopf algebra
and~$\repX$ an $A$-module.  The balancing element $\balance$ allows
constructing the ``canonical'' $q$-character of~$\repX$:
\begin{equation}\label{qCh}
  \qTr_{\repX}\equiv\qtr^\balance_{\repX}=\tr_{\repX}(\balance^{-1}?)
  \in\Ch(A).
\end{equation}

We also define the quantum dimension of a module $\repX$ as
\begin{gather*}
  \qdim\repX= \Tr_{\repX}\balance^{-1}.
\end{gather*}
It satisfies the relation
\begin{equation*}
  \qdim\repX_1\tensor\repX_2=\qdim\repX_1\qdim\repX_2.
\end{equation*}
for any two modules $\repX_1$ and $\repX_2$.

Let now $A$ be a factorizable ribbon Hopf algebra and let $\Grring(A)$
be its Grothendieck ring.  We combine the mapping $\Grring(A)\to A^*$
given by $\repX\mapsto\qTr_{\repX}$ and the Drinfeld mapping
$\drmap:A^*\to A$.
\begin{lemma}\label{lemma:Dr-hom}
  In a factorizable ribbon Hopf algebra~$A$, the mapping
  \begin{equation*}
    \drmap\circ\qTr:
    \Grring(A)
    \to
    \cZ(A)
  \end{equation*}
  is a homomorphism of associative commutative algebras.
\end{lemma}

\section{The quantum double}\label{sec:double}
In this Appendix, we construct a double of the Hopf algebra $B$
associated with the short screening in the logarithmic conformal field
theory outlined in~\bref{sec:VOA}.  The main structure resulting from
the double is the $R$-matrix, which is then used to construct the
$M$-matrix $\bar M$ for~$\UresSL2$.

\subsection{Constructing a double of the ``short-screening'' quantum
  group}\label{subsec:double} For $\q\,{=}\,e^{\frac{i\pi}{p}}$, we
let $B$ denote the Hopf algebra generated by~$\dE$ and~$\dK$ with the
relations
\begin{equation}\label{Hopf-start}
  \begin{gathered}
   \dE^p=0,\quad\dK^{4p}=\one,\quad\dK\dE\dK^{-1}=\q\dE,\\
    \Delta(\dE)=\one\otimes\dE+\dE\otimes\dK^2,\quad
    \Delta(\dK)=\dK\otimes\dK,\\
    \epsilon(\dE)=0,\quad\epsilon(\dK)=1,\\
    S(\dE)=-\dE\dK^{-2},\quad S(\dK)=\dK^{-1}.
  \end{gathered}
\end{equation}
The PBW-basis in $B$ is
\begin{equation*}
  \pbw_{mn}=\dE^m\dK^n,\quad 0\leq m\leq p-1,\quad 0\leq n\leq 4p-1.
\end{equation*}

The space $B^*$ of linear functions on $B$ is a Hopf algebra with the
multiplication, comultiplication, unit, counit, and antipode given by
\begin{equation}\label{double-def}
  \begin{gathered}
    \coup{\beta\gamma}{x}=\sum_{(x)}\coup{\beta}{x'}
    \coup{\gamma}{x''},\quad
    \coup{\Delta(\beta)}{x\tensor y}=\coup{\beta}{yx},\\
    \coup{\one}{x}=\epsilon(x),\quad
    \epsilon(\beta)=\coup{\beta}{\one},\quad
    \coup{S(\beta)}{x}=\coup{\beta}{S^{-1}(x)}
  \end{gathered}
\end{equation}
for any $\beta,\gamma\,{\in}\, B^*$ and $x,y\,{\in}\, B$.  

The quantum double $D(B)$ is a Hopf algebra with the underlying vector
space $B^*\tensor B$ and with the multiplication, comultiplication,
unit, counit, and antipode given by Eqs.~\eqref{Hopf-start}
and~\eqref{double-def} and~by
\begin{equation}\label{double-def-1}
  x\beta=\sum_{(x)}\beta(S^{-1}(x''')?x')x'',
  \qquad x\in B,\quad\beta\in B^*.
\end{equation}

\begin{thm}\label{thm:double}  
  $D(B)$ is the Hopf algebra generated by $\dE$, $\dF$, $\dK$, and
  $\ddK$ with the relations
  \begin{gather}
    \dK\dE\dK^{-1}=\q\dE,\quad\dE^p=0,\quad
    \dK^{4p}=\one,\label{rel-B-1}\\
    \ddK\dF\ddK^{-1}=\q\dF,\quad\dF^p=0,\quad
    \ddK^{4p}=\one,\label{rel-B*-1}\\
    \dK\ddK=\ddK\dK,\quad\dK\dF\dK^{-1}=\q^{-1}\dF,\quad
    \ddK\dE\ddK^{-1}=\q^{-1}\dE,\quad
    [\dE,\dF]=\ffrac{\dK^2-\ddK^2}{\q-\q^{-1}},\label{rel-BB*}\\
    \Delta(\dE)=\one\otimes\dE+\dE\otimes\dK^2,\quad
    \Delta(\dK)=\dK\otimes\dK,\quad
    \epsilon(\dE)=0,\quad
    \epsilon(\dK)=1,\label{rel-B-2}\\
    \Delta(\dF)=\ddK^2\tensor\dF+\dF\tensor\one,\quad
    \Delta(\ddK)=\ddK\tensor\ddK,\quad
    \epsilon(\dF)=0,\quad
    \epsilon(\ddK)=1,\label{rel-B*-2}\\
    S(\dE)=-\dE\dK^{-2},\quad
    S(\dK)=\dK^{-1},\label{rel-B-3}\\
    S(\dF)=-\ddK^{-2}\dF,\quad S(\ddK)=\ddK^{-1}.\label{rel-B*-3}
  \end{gather}
\end{thm}
\begin{proof}
  Equations~\eqref{rel-B-1}, \eqref{rel-B-2}, and \eqref{rel-B-3} are
  relations in $B$.  The unit in $B^*$ is given by the function $\one$
  such that
  \begin{equation*}
    \coup{\one}{\pbw_{mn}}=\delta_{m,0}.
  \end{equation*}
  The elements $\ddK,\dF\,{\in}\, B^*$ are uniquely defined by
  \begin{equation*}
    \coup{\ddK}{\pbw_{mn}}=\delta_{m,0}\q^{-n/2},\qquad
    \coup{\dF}{\pbw_{mn}}=\delta_{m,1}\ffrac{\q^{-n}}{\q-\q^{-1}}.
  \end{equation*}
  
  For elements of the PBW-basis of $B$, the first relation
  in~\eqref{double-def} becomes
  \begin{equation}\label{multiplic}
    \coup{\beta\gamma}{\pbw_{mn}}
    =\sum_{r=0}^m\qbinom{m}{r}\coup{\beta}{\dE^{m-r}\dK^n}
    \coup{\gamma}{\dE^{r}\dK^{2m-2r+n}},
  \end{equation}
  where we use the notation
  \begin{equation*}
    \angint{n} = \ffrac{q^{2n}-1}{q^2-1}=q^{n-1}[n],\quad
    \angint{n}! = \angint{1}\angint{2}\dots
    \angint{n},\quad
    \qbinom{m}{n}=\ffrac{\angint{m}!}{\angint{n}!\,\angint{m-n}!}.
  \end{equation*}
  We then check that the elements
  $\dF^i\ddK^j$ with $0\leq i\leq p-1$ and $0\leq j\leq 4p-1$
  constitute a basis in $B^*$ and evaluate on the basis elements of
  $B$ as
  \begin{equation}\label{basis}
    \coup{\dF^i\ddK^j}{\pbw_{mn}}
    =\delta_{mi}\ffrac{\angint{i}!}{
      (\q-\q^{-1})^i}\q^{-(j+2i)n/2-ij -i(i-1)}
  \end{equation}
  The easiest way to see that~\eqref{basis} holds is to
  use~\eqref{multiplic} to calculate $\coup{\dF^j}{\dE^m\dK^n}$ and
  $\coup{\ddK^j}{\dE^m\dK^n}$ by induction on $j$ and then calculate
  $\coup{\dF^i\ddK^j}{\dE^m\dK^n}$ using~\eqref{multiplic} again, with
  $\beta=\dF^i$ and $\gamma=\ddK^j$.
  
  Next, we must show that $\dF^i\ddK^j$ are linearly independent for
  $0\leq i\leq p-1$ and $0\leq j\leq$\linebreak[0]$4p-1$.  Possible
  linear dependences are
  $\sum_{i=0}^{p-1}\sum_{j=0}^{4p-1}\lambda_{ij}\dF^i\ddK^j=0$ with
  some $\lambda_{ij}\,{\in}\,\oC$, that~is,
  \begin{equation*}
    \sum_{i=0}^{p-1}\sum_{j=0}^{4p-1}
    \lambda_{ij}\coup{\dF^i\ddK^j}{\dE^m\dK^n}=0
  \end{equation*}
  for all $0\leq m\leq p-1$ and $0\leq n\leq 4p-1$.  Using
  \eqref{basis}, we obtain the system of $4p^2$ linear equations
  \begin{multline*}
    \smash[b]{\sum_{i=0}^{p-1}\sum_{j=0}^{4p-1}
      \delta_{mi}\ffrac{\angint{i}!}{(\q-\q^{-1})^i}}\,\q^{-(j+2i)n/2-ij
      -i(i-1)}
    \lambda_{ij}={}\\
    {}=\ffrac{\angint{m}!}{ (\q-\q^{-1})^m}\,
    \q^{-mn -m(m-1)}\sum_{j=0}^{4p-1} \q^{-\frac{1}{2} j(n+2m)}
    \lambda_{mj}=0
  \end{multline*}
  for the $4p^2$ variables $\lambda_{ij}$.  The system decomposes into
  $p$ independent systems of $4p$ linear equations
  \begin{equation*}
    \sum_{j=0}^{4p-1} A_{jn} \lambda_{mj}=0
  \end{equation*}
  for $4p$ variables $\lambda_{mj}$, $0\leq j\leq 4p-1$ (with $m$
  fixed), where $A_{jn}=\q^{-\frac{1}{2}j(n+2m)}$.  The determinant of
  the matrix $A_{jn}$ is the Vandermonde determinant, which is nonzero
  because no two numbers among $(\q^{-\frac{1}{2}(n+2m)})_{0\leq n\leq
    4p-1}$ coincide.
      
  With~\eqref{basis} established, we verify \eqref{rel-B*-1},
  \eqref{rel-B*-2}, and~\eqref{rel-B*-3}.
  
  Next, to verify \eqref{rel-BB*}, we write~\eqref{double-def-1} for
  $x=\dK$ and $x=\dE$ as the respective relations
  \begin{equation}\label{commut}
    \dK\beta=\beta(\dK^{-1}?\dK)\dK,\qquad\dE\beta
    =-\beta(\dK^{-2}\dE?)+\beta(\dK^{-2}?)\dE
    +\beta(\dK^{-2}?\dE)\dK^2
  \end{equation}  
  valid for all $\beta\,{\in}\, B^*$.  The following formulas are
  obtained by direct calculation using~\eqref{basis}:
  \begin{alignat*}{2}
    \ddK(\dK^{-1}?\dK)&=\ddK,&\qquad\ddK(\dK^{-2}\dE?)&=0,\\
    \ddK(\dK^{-2}?)&=\q\ddK,&\qquad\ddK(\dK^{-2}?\dE)&=0,\\
    \dF(\dK^{-1}?\dK)&=\q^{-1}\dF,&\qquad
    \dF(\dK^{-2}\dE?)&=\ffrac{\ddK^2}{\q-\q^{-1}},\\
    \dF(\dK^{-2}?)&=\dF,&\qquad\dF(\dK^{-2}?\dE)
    &=\ffrac{\one}{\q-\q^{-1}}.
  \end{alignat*}
  These relations and \eqref{commut} imply \eqref{rel-BB*}, which
  finishes the proof.
\end{proof}

\subsection{The $R$-matrix} As any Drinfeld double, $D(B)$ is a
quasitriangular Hopf algebra, with the universal $R$-matrix given by
\begin{equation}\label{q-double-R}
  R=\sum_{m=0}^{p-1}\sum_{i=0}^{4p-1}\pbw_{mi}\tensor\dpbw_{mi},
\end{equation}
where $\pbw_{mi}$ are elements of a basis in~$B$ and
$\dpbw_{ij}\,{\in}\, B^*$ are elements of the dual basis,
\begin{equation}\label{dual}
  \coup{\dpbw_{ij}}{\pbw_{mn}}=\delta_{im}\delta_{jn}.
\end{equation}

\begin{lemma}\label{lemma:R}
  For $D(B)$ constructed in~\bref{subsec:double}, the dual basis is
  expressed in terms of the generators $\dF$ and $\ddK$ as
  \begin{equation}\label{dpbw}
    \dpbw_{ij}=
    \ffrac{(\q-\q^{-1})^i}{[i]!}\,\q^{i(i-1)/2}
    \ffrac{1}{4p}\sum_{r=0}^{4p-1}
    \q^{i(j+r)+rj/2}\dF^i\ddK^r,
  \end{equation}
  and therefore the $R$-matrix is given by
  \begin{equation}\label{the-R}
    R =\ffrac{1}{4p}\sum_{m=0}^{p-1}\sum_{i,j=0}^{4p-1}
    \ffrac{(\q-\q^{-1})^m}{[m]!}\,\q^{m(m-1)/2+m(i-j)-ij/2}
    \dE^m\dK^{i}\otimes\dF^m\ddK^{-j}.
  \end{equation}  
\end{lemma}
\begin{proof}
  By a direct calculation using \eqref{basis}, we verify that
  Eqs.~\eqref{dual} are satisfied with $\dpbw_{ij}$ given
  by~\eqref{dpbw}.
\end{proof}

\section{Verma and projective modules}\label{verma-proj-mod-base}

\subsection{Verma and contragredient Verma
  modules}\label{verma-mod-base} Let $s$ be an integer $1\leq s\leq
p-1$ and $\alpha=\pm1$.  The Verma module $\Verma^{\alpha}(s)$ has the
basis
\begin{equation}\label{app:Verma-basis}
  \{\leftpr_k\}_{0\le k\le s-1}
  \cup\{\botpr_n\}_{0\le n\le p-s-1},
\end{equation}
where $\{\botpr_n\}_{0\le n\le p-s-1}$ correspond to the submodule
$\repX^{-\alpha}(p-s)$ and $\{\leftpr_k\}_{0\le k\le s-1}$ correspond
to the quotient module $\repX^{\alpha}(s)$
in
\begin{equation}\label{verma-ext}
  0\to\repX^{-\alpha}(p-s)\to
  \Verma^{\alpha}(s)\to\repX^{\alpha}(s)\to0,
\end{equation}
with the $\UresSL2$-action given by
\begin{alignat}{3}\label{app:Verma-EK-action}
  K\leftpr_k&=\alpha \q^{s-1-2k}\leftpr_k,& \quad &0\le k\le s-1,\notag\\
  K\botpr_n&=-\alpha \q^{p-s-1-2n}\botpr_n,& \quad &0\le n\le p-s-1,
  \notag\\
  E\leftpr_k&=\alpha [k][s-k]\leftpr_{k-1},& \quad &0\le k\le s-1
  \quad(\text{with}\quad\leftpr_{-1}\equiv0),
  \kern-60pt
\end{alignat}
\begin{align}\label{app:Verma-F-action}
  E\botpr_n&=-\alpha [n][p-s-n]\botpr_{n-1},
  \quad 0\le n\le p-s-1\quad(\text{with}\quad\botpr_{-1}\equiv0)\notag\\
  \intertext{and}
  F\leftpr_k&=
  \begin{cases}
    \leftpr_{k+1}, &0\le k\le s-2,\\
    \botpr_0, & k=s-1,\\
  \end{cases}
  \\
  F\botpr_n&=\botpr_{n+1}, \quad 0\le n\le p-s-1
  \quad(\text{with}\quad\botpr_{p-s}\equiv0).\notag
\end{align}

In addition, there are Verma modules $\Verma^{\pm}(p)=\repX^{\pm}(p)$.

The contragredient Verma module $\CVerma^{\alpha}(s)$ is defined in
the basis \eqref{app:Verma-basis} by the same formulas except
\eqref{app:Verma-EK-action} and \eqref{app:Verma-F-action}, replaced
by the respective formulas
\begin{align*}
  E\leftpr_k&=
  \begin{cases}
    \botpr_{p-s-1}, & k=0,\\
    \alpha [k][s-k]\leftpr_{k-1}, &1\le k\le s-1,\\
  \end{cases}
  \\
  F\leftpr_k&=\leftpr_{k+1}, \quad 0\le k\le s-1
  \quad(\text{with}\quad\leftpr_{s}\equiv0).
\end{align*}

\subsection{Projective modules}\label{proj-mod-base}
The module $\mathscr{P}^{\pm}(s)$, $1\leq s\leq p-1$, is the
projective module whose irreducible quotient is given
by~$\repX^{\pm}(s)$.  The modules $\mathscr{P}^{\pm}(s)$ appeared in
the literature several times, see~\cite{[RT],[GL],[JMT]}.  In
explicitly describing their structure, we follow~\cite{[JMT]} most
closely.

\subsubsection{$\boldsymbol{\modL(s)}$}\label{module-L}
Let $s$ be an integer $1\leq s\leq p-1$.  The projective module
$\modL(s)$ has the basis
\begin{equation*}
  \{\leftpr^{(+,s)}_k,\rightpr^{(+,s)}_k\}_{0\le k\le p-s-1}
  \cup\{\botpr^{(+,s)}_n,\toppr^{(+,s)}_n\}_{0\le n\le s-1},
\end{equation*}
where $\{\toppr^{(+,s)}_n\}_{0\le n\le s-1}$ is the basis
corresponding to the top module in~\eqref{schem-proj},\\
$\{\botpr^{(+,s)}_n\}_{0\le n\le s-1}$ to the bottom ,
$\{\leftpr^{(+,s)}_k\}_{0\le k\le p-s-1}$ to the left, and
$\{\rightpr^{(+,s)}_k\}_{0\le k\le p-s-1}$ to the right module, with
the $\UresSL2$-action given by
\begin{alignat*}{3}
  K\leftpr^{(+,s)}_k&=-\q^{p-s-1-2k}\leftpr^{(+,s)}_k,& \quad
  K\rightpr^{(+,s)}_k&=-\q^{p-s-1-2k}\rightpr^{(+,s)}_k,&
  \quad &0\le k\le p-s-1,\\
  K\botpr^{(+,s)}_n&=\q^{s-1-2n}\botpr^{(+,s)}_n,& \quad
  K\toppr^{(+,s)}_n&=\q^{s-1-2n}\toppr^{(+,s)}_n,& \quad &0\le n\le
  s-1,\\
  E\leftpr^{(+,s)}_k&=-[k][p-s-k]\leftpr^{(+,s)}_{k-1},&
  \quad 0\le k&\le p-s-1
  \quad(\text{with}\quad\leftpr^{(+,s)}_{-1}\equiv0),
  \kern-60pt
\end{alignat*}
\begin{align*}
  E\rightpr^{(+,s)}_k&=
  \begin{cases}
    -[k][p-s-k]\rightpr^{(+,s)}_{k-1}, &1\le k\le p-s-1,\\
    \botpr^{(+,s)}_{s-1}, & k=0,\\
  \end{cases}
  \\
  E\botpr^{(+,s)}_n&=[n][s-n]\botpr^{(+,s)}_{n-1},
  \quad 0\le n\le s-1\quad(\text{with}\quad\botpr^{(+,s)}_{-1}\equiv0),\\
  E\toppr^{(+,s)}_n&=
  \begin{cases}
    [n][s-n]\toppr^{(+,s)}_{n-1}+\botpr^{(+,s)}_{n-1}, &1\le n\le s-1,\\
    \leftpr^{(+,s)}_{p-s-1}, & n=0,\\
  \end{cases}
  \\
  \intertext{and}
  F\leftpr^{(+,s)}_k&=
  \begin{cases}
    \leftpr^{(+,s)}_{k+1}, &0\le k\le p-s-2,\\
    \botpr^{(+,s)}_0, & k=p-s-1,\\
  \end{cases}
  \\
  F\rightpr^{(+,s)}_k&=\rightpr^{(+,s)}_{k+1}, \quad 0\le k\le p-s-1
  \quad(\text{with}\quad\rightpr^{(+,s)}_{p-s}\equiv0),\\
  F\botpr^{(+,s)}_n&=\botpr^{(+,s)}_{n+1}, \quad 0\le n\le s-1
  \quad(\text{with}\quad\botpr^{(+,s)}_s\equiv0),\\
    F\toppr^{(+,s)}_n&=
  \begin{cases}
    \toppr^{(+,s)}_{n+1}, &0\le n\le s-2,\\
    \rightpr^{(+,s)}_0, & n=s-1.
  \end{cases}
\end{align*}

\subsubsection{$\boldsymbol{\modP(p-s)}$}\label{module-P}
Let $s$ be an integer $1\leq s\leq p-1$.  The projective module
$\modP(p-s)$ has the basis
\begin{equation*}
  \{\leftpr^{(-,s)}_k,\rightpr^{(-,s)}_k\}_{0\le k\le p-s-1}
  \cup\{\botpr^{(-,s)}_n,\toppr^{(-,s)}_n\}_{0\le n\le s-1},
\end{equation*}
where $\{\rightpr^{(-,s)}_k\}_{0\le k\le p-s-1}$ is the basis
corresponding to the top module in~\eqref{schem-proj},\\
$\{\leftpr^{(-,s)}_k\}_{0\le k\le p-s-1}$ to the bottom,
$\{\botpr^{(-,s)}_n\}_{0\le n\le s-1}$ to the left, and
$\{\toppr^{(-,s)}_n\}_{0\le n\le s-1}$ to the right module, with the
$\UresSL2$-action given by
\begin{alignat*}{3}
  K\leftpr^{(-,s)}_k&=-\q^{p-s-1-2k}\leftpr^{(-,s)}_k,& \quad
  K\rightpr^{(-,s)}_k&=-\q^{p-s-1-2k}\rightpr^{(-,s)}_k,&
  \quad &0\le k\le p-s-1,\\
  K\botpr^{(-,s)}_n&=\q^{s-1-2n}\botpr^{(-,s)}_n,& \quad
  K\toppr^{(-,s)}_n&=\q^{s-1-2n}\toppr^{(-,s)}_n,& \quad &0\le n\le
  s-1,\\
  E\leftpr^{(-,s)}_k&=-[k][p-s-k]\leftpr^{(-,s)}_{k-1},& \quad
  0\le k&\le p-s-1\quad(\text{with}\quad\leftpr^{(-,s)}_{-1}\equiv0),
  \kern-60pt
\end{alignat*}
\begin{align*}
  E\rightpr^{(-,s)}_k&=
  \begin{cases}
    -[k][p-s-k]\rightpr^{(-,s)}_{k-1}+\leftpr^{(-,s)}_{k-1},
    &1\le k\le p-s-1,\\
    \botpr^{(-,s)}_{s-1}, & k=0,\\
  \end{cases}
  \\
  E\botpr^{(-,s)}_n&=[n][s-n]\botpr^{(-,s)}_{n-1},
  \quad 0\le n\le s-1\quad(\text{with}\quad
  \botpr^{(-,s)}_{-1}\equiv0),\\
  E\toppr^{(-,s)}_n&=
  \begin{cases}
    [n][s-n]\toppr^{(-,s)}_{n-1}, &1\le n\le s-1,\\
    \leftpr^{(-,s)}_{p-s-1}, & n=0,\\
  \end{cases}
  \\
  \intertext{and}
  F\leftpr^{(-,s)}_k&=\leftpr^{(-,s)}_{k+1}, \quad 0\le k\le p-s-1
  \quad(\text{with}\quad\leftpr^{(-,s)}_{p-s}\equiv0),\\
  F\rightpr^{(-,s)}_k&=
  \begin{cases}
    \rightpr^{(-,s)}_{k+1}, &0\le k\le p-s-2,\\
    \toppr^{(-,s)}_0, & k=p-s-1,\\
  \end{cases}
  \\
  F\botpr^{(-,s)}_n&=
  \begin{cases}
    \botpr^{(-,s)}_{n+1}, &0\le n\le s-2,\\
    \leftpr^{(-,s)}_0, & n=s-1,
  \end{cases}
  \\
  F\toppr^{(-,s)}_n&=\toppr^{(-,s)}_{n+1}, \quad 0\le n\le s-1
  \quad(\text{with}\quad\toppr^{(-,s)}_s\equiv0).
\end{align*}

\section{Construction of the canonical central
  elements}\label{app:center}

\subsection{Canonical central elements}
To explicitly construct the canonical central elements
in~\bref{prop-center} in terms of the $\UresSL2$ generators, we use
the standard formulas in~\cite[Ch.~V.2]{[Gant]} (also
cf.~\cite{[Kerler]}; we are somewhat more explicit about the
representation-theory side, based on the analysis
in~\bref{the-center}).  We first introduce projectors $\pi^+_s$ and
$\pi^-_s$ on the direct sums of the eigenspaces of~$K$ appearing in
the respective representations~$\repLambda(s)$ and~$\repPi(p-s)$ for
$1\leq s\leq p-1$, Eqs.~\eqref{eq:K-eigens-Lambda}
and~\eqref{eq:K-eigens-Pi}.  These projectors are
\begin{equation}\label{projectors}
    \pi^+_s=\ffrac{1}{2p}\sum_{n=0}^{s-1}
    \sum_{j=0}^{2p-1}\q^{(2n-s+1)j}K^j
    ,\qquad
    \pi^-_s= \ffrac{1}{2p}\sum_{n=s}^{p-1}
    \sum_{j=0}^{2p-1}\q^{(2n-s+1)j}K^j.
\end{equation}
It follows that
\begin{equation}\label{pi+pi}
  \pi^+_s+\pi^-_s=\half(\one-(-1)^{s}K^p).
\end{equation}

Second, we recall polynomial relation~\eqref{Cas-relation} for the
 Casimir element and define the polynomials
 \begin{multline*}
   \psi_0(x)=(x-\beta_p)\prod_{r=1}^{p-1}(x-\beta_r)^2,
   \\
   \psi_s(x)=(x-\beta_0)\,(x-\beta_p)
   \smash{\prod_{\substack{r=1\\
         r\neq s}}^{p-1}}(x-\beta_r)^2,\quad 1\leq s\leq p-1,\\
   \psi_p(x)=(x-\beta_0)\prod_{r=1}^{p-1}(x-\beta_r)^2,
\end{multline*}
where we recall that $\beta_j=\ffrac{\q^j+\q^{-j}}{(\q-\q^{-1})^2}$,
with $\beta_j\neq\beta_{j'}$ for $0\leq j\neq j'\leq p$.

\begin{prop}\label{prop-center-explicit}
  The canonical central elements $\idem_s$, $0\leq s\leq p$, and
  $\nilp_s$, $1\leq s\leq p-1$, are explicitly given as follows.  The
  elements in the radical of~$\cZ$ are
  \begin{equation}\label{center-rad}
    \nilp^{\pm}_s=\pi^{\pm}_s \nilp_s,\qquad
    1\leq s\leq p-1,
  \end{equation}
  where
  \begin{gather}\label{center-Rad}
    \nilp_s = \ffrac{1}{\psi_s(\beta_s)}
    \bigl(\cas-\beta_s\bigr)\psi_s(\cas).
  \end{gather}
  The canonical central idempotents are given by  
  \begin{equation}\label{center-proj}
    \idem_s=\ffrac{1}{\psi_s(\beta_s)}
    \bigl(\psi_s(\cas) - \psi'_s(\beta_s)\nilp_s\bigr),
    \quad 0\leq s\leq p,
  \end{equation}
  where we formally set $\nilp_0=\nilp_p=0$.  
\end{prop}

\begin{proof}
  First, $(\cas-\beta_r)\psi_r(\cas)$ acts by zero on
  $\modQ(0)=\repPi(p)\boxtimes\repPi(p)$ and
  $\modQ(p)=\repLambda(p)\boxtimes\repLambda(p)$.  We next consider
  its action on~$\modQ(s)$ for $1\leq s\leq p-1$.  It follows
  from~\bref{proj-mod-base} that the Casimir element acts on the basis
  of $\modL(s)$ as
  \begin{equation}\label{cas-action}
    \begin{gathered}
      \cas\toppr^{(+,s)}_n=\beta_s \toppr^{(+,s)}_n+\botpr^{(+,s)}_n,\\*
      \cas\leftpr^{(+,s)}_n =\beta_s\leftpr^{(+,s)}_n,\qquad
      \cas\rightpr^{(+,s)}_n=\beta_s \rightpr^{(+,s)}_n,\\*
      \cas\botpr^{(+,s)}_n =\beta_s\botpr^{(+,s)}_n
    \end{gathered}
  \end{equation}
  for all $0\le n\le s-1$.  Clearly, $(\cas-\beta_s)^2$ annihilates
  the entire $\modL(s)$, and therefore $(\cas-\beta_r)\psi_r(\cas)$
  acts by zero on each $\modQ(s)$ with $s\neq r$.  On the other hand,
  for $s=r$, we have\pagebreak[3]
  \begin{equation*}
    (\cas-\beta_r)\psi_r(\cas)\toppr^{(+,r)}_n=
    \psi_r(\cas)\botpr^{(+,r)}_n=\psi_r(\beta_r)\botpr^{(+,r)}_n.
  \end{equation*}
  Similar formulas describe the action of the Casimir element on the
  module $\modP(p-s)$.  It thus follows that $\nilp_r$ sends the
  quotient of the bimodule $\modQ(r)$ in~\eqref{prop:subquotient:Q/R},
  i.e., $\repLambda(r)\boxtimes\repLambda(r)\oplus
  \repPi(p-r)\boxtimes\repPi(p-r)$, into the subbimodule
  $\repLambda(r)\boxtimes\repLambda(r)\oplus
  \repPi(p-r)\boxtimes\repPi(p-r)$ at the bottom of $\modQ(r)$.
  Therefore,~$\nilp_r=\mathrm{const}\cdot(\nilp^+_r+\nilp^-_r)$.
    
  To obtain~$\nilp^+_r$ and~$\nilp^-_r$, we multiply $\nilp_r$ with
  the respective operators projecting on the direct sums of the
  eigenspaces of $K$ occurring in~$\repLambda(s)$ and~$\repPi(p-s)$.
  This gives~\eqref{center-rad} (the reader may verify independently
  that although the projectors $\pi^\pm_r$ are not central, their
  products with $\nilp_r$ are).  The normalization
  in~\eqref{center-Rad} is chosen such that we have
  $\nilp_r\toppr^{(+,r)}_n=\botpr^{(+,r)}_n$.
  
  To obtain the idempotents $\idem_r$, we note that
  $\psi_r(\cas)$ annihilates all $\modQ(s)$ for $s\neq r$,
  while on $\modQ(r)$, we have
  $\psi_r(\cas)\leftpr^{(+,r)}_n
  =\psi_r(\beta_r)\leftpr^{(+,r)}_n$, $\psi_r(\cas)\rightpr^{(+,r)}_n
  =\psi_r(\beta_r)\rightpr^{(+,r)}_n$, $\psi_r(\cas)\botpr^{(+,r)}_n
  =\psi_r(\beta_r)\botpr^{(+,r)}_n$, and furthermore, by Taylor
  expanding the polynomial, 
  \begin{equation*}
    \psi_r(\cas)\toppr^{(+,r)}_n =\psi_r(\beta_r)\toppr^{(+,r)}_n +
    (\cas-\beta_r)\psi_r'(\beta_r)\toppr^{(+,r)}_n,
  \end{equation*}
  with higher-order terms in $(\cas-\beta_r)$ annihilating
  $\toppr^{(+,r)}_n$.  Similar formulas hold for the action
  on~$\modP(p-s)$.  Therefore, $\modQ(r)$ is the root space of
  $\frac{1}{\psi_r(\beta_r)}\psi_r(\cas)$ with eigenvalue~$1$, and the
  second term in~\eqref{center-proj} is precisely the subtraction of
  the nondiagonal part.
\end{proof}

\subsection{Remarks}\label{remarks-D}\mbox{}

\begin{enumerate}
  
\item\label{item:w} We note that $\nilp^+_s + \nilp^-_s=\nilp_s$.
  This follows because $\bigl(1+(-1)^s K^p\bigr)\nilp_s=0$.
  
\item For any polynomial~$\polR(\cas)$,\ 
  decomposition~\eqref{decomp-general} takes the form
  \begin{equation}\label{P(C)-idem}
    \polR(\cas)=\sum_{s=0}^{p}
    \polR(\beta_s) \idem_s + \sum_{s=1}^{p-1}\polR'(\beta_s)\nilp_s.
  \end{equation}
  For example,~\eqref{P(C)-idem} implies that for $\hat\cas$ defined
  in~\bref{sec:Casimir}, we have
  \begin{equation*}
    \hat\cas
    = \sum_{s=0}^{p} (\q^s + \q^{-s}) \idem_{s}
    + (\q - \q^{-1})^2 \sum_{s=1}^{p - 1} \nilp_{s}.
  \end{equation*}
\end{enumerate}

\subsection{Eigenmatrix of the $(1,p)$ fusion algebra}\label{eigenP}
Using~\eqref{P(C)-idem} and expressions through the Chebyshev
polynomials in~\bref{prop:quotient}, we recover the eigenmatrix
$\eigenP$ of the fusion algebra~\eqref{the-fusion}.  This eigenmatrix
was obtained in~\cite{[FHST]} by different means, from the matrix of
the modular $S$-transformation on $\algW(p)$-characters.  The
eigenmatrix relates the preferred basis (the basis of irreducible
representations) and the basis of idempotents and nilpotents in the
fusion algebra.  Specifically, if we order the irreducible
representations as
\begin{equation*}
  \mat{X}^t\equiv(\repX^+(p),
  \repX^-(p), \repX^+(1),\repX^-(p-1),\dots, \repX^+(p-1),\repX^-(1))
\end{equation*}
and the idempotents and nilpotents that form a basis of
$\Drinalg_{2p}\cong\Grring_{2p}$ as
\begin{equation*}
  \mat{Y}^t\equiv(\idem_p,\idem_0,\idem_1,\nilp_1,\dots,
  \idem_{p-1},\nilp_{p-1}), 
\end{equation*}
then the eigenmatrix $\eigenP(p)$ is defined as
\begin{equation*}
  \mat{X}=\eigenP(p)\,\mat{Y}.
\end{equation*}
The calculation of the entries of~$\eigenP(p)$ via~\eqref{P(C)-idem}
is remarkably simple: for example, with $\polR(\hat\cas)$ taken
as~$\cheb_s(\hat\cas)$ (see~\bref{prop:quotient}), we have
\begin{equation*}
  \polR(\hat\beta_j)=\polR(2\cos\ffrac{\pi
    j}{p}) =\mfrac{\sin\frac{\pi j s}{p}}{\sin\frac{\pi j}{p}}
\end{equation*}
in accordance with~\eqref{eq:chebyshev-sin}.  Evaluating the other
case in~\eqref{basis-P} similarly and taking the derivatives, we
obtain the eigenmatrix
\begin{equation*}
  \eigenP(p)=
  \begin{pmatrix}
    P_{0,0}&P_{0,1}&\dots&P_{0,p-1}\\
    P_{1,0}&P_{1,1}&\dots&P_{1,p-1}\\
    \vdots&\vdots&\ddots&\vdots\\
    P_{p-1,0}&P_{p-1,1}&\dots&P_{p-1,p-1}
  \end{pmatrix}
\end{equation*}
with the $2\,{\times}\,2$ blocks~\cite{[FHST]}\footnote{The formula
  for $P_{0,j}$ corrects a misprint in~\cite{[FHST]}, where
  $(-1)^{j+p}$ occurred in a wrong matrix entry.}
\begin{multline*}
  \begin{alignedat}{2}
    P_{0,0}&=
    \begin{pmatrix}
      p\; & (-1)^{p+1}p\\
      p\; & -p
    \end{pmatrix}\!,
    &\quad P_{0,j}&=
    \begin{pmatrix}
      0\;&-(-1)^{j+p}\,\ffrac{2\lambda_j}{p}\sin\ffrac{j\pi}{p}\\[9pt]
      0\;&-\ffrac{2\lambda_j}{p}\sin\ffrac{j\pi}{p}
    \end{pmatrix}\!,
    \\
    P_{s,0}&=
    \begin{pmatrix}
      s\; & (-1)^{s+1}s\\[2pt]
      p{-}s\; & (-1)^{s+1}(p{-}s)
    \end{pmatrix}\!,
  \end{alignedat}\\
  P_{s,j}=
  (-1)^s
  \begin{pmatrix}
    -\ffrac{\sin\frac{sj\pi}{p}}{\sin\frac{j\pi}{p}}
    & \ffrac{2\lambda_j}{p^2}
    \Bigl(-s\cos\ffrac{sj\pi}{p}\sin\ffrac{j\pi}{p}
    +\sin\ffrac{sj\pi}{p}\cos\ffrac{j\pi}{p}\Bigr)\\[12pt]
    \ffrac{\sin\frac{sj\pi}{p}}{\sin\frac{j\pi}{p}}
    &\ffrac{2\lambda_j}{p^2}\Bigl(
    -(p{-}s)\cos\ffrac{sj\pi}{p}\sin\ffrac{j\pi}{p}
    -\sin\ffrac{sj\pi}{p}\cos\ffrac{j\pi}{p}\Bigr)
  \end{pmatrix}
\end{multline*}
for $s,j\,{=}\,1,\dots,p{-}1$, where, for the sake of comparison, we
isolated the factor
\begin{equation*}
  \lambda_j=\mfrac{p^2}{[j]^3\sin\frac{\pi}{p}}
  =\mfrac{p^2\,\bigl(\sin\frac{\pi}{p}\bigr)^2}{
    \bigl(\sin\frac{j\pi}{p}\bigr)^3}
\end{equation*}
whereby the normalization of each nilpotent element, and hence of each
even column of~$\eigenP$ starting with the fourth, differs from the
normalization chosen in~\cite{[FHST]} (both are arbitrary because the
nilpotents cannot be canonically normalized).

\section{Derivation of the $q$-binomial identity}\label{app:derivation}
We derive identity~\eqref{the-identity} from the fusion algebra
realized on the central elements $\cchi^{\pm}(s)$.  In view
of~\bref{lemma:Dr-hom}, the central elements $\cchi^{\alpha}(s)$
in~\eqref{the-cchi} \textup{(}with $\alpha=\pm1$,
$s=1,\dots,p$\textup{)} satisfy the algebra\pagebreak[3]
\begin{equation}\label{the-fusion-cchi}
  \cchi^{\alpha}(s)\,\cchi^{\alpha'}(s')
  =\sum_{\substack{s''=|s - s'| + 1\\
      \mathrm{step}=2}}^{s + s' - 1}
  \widetilde\cchi^{\alpha\alpha'}(s''),
\end{equation}
where
\begin{equation*}
  \widetilde\cchi^{\alpha}(s)
  =
  \begin{cases}
    \cchi^{\alpha}(s),&1\leq s\leq p,\\
    \cchi^{\alpha}( 2p - s) + 2\cchi^{-\alpha}( s - p),
    & p + 1 \leq s \leq 2p - 1.
  \end{cases}
\end{equation*}

We now equate the coefficients at the respective PBW-basis elements in
both sides of~\eqref{the-fusion-cchi}.  Because
of~\eqref{minus-alpha}, it suffices to do this for the algebra
relation for~$\cchi^{+}(s)\,\cchi^{+}(s')$.  Writing it as
in~\eqref{rewrite_v2}, we have
\begin{multline}\label{lambda-id}
  \cchi^{+}(s)\,\cchi^+(s')
  = \sum_{\substack{s''=|s - s'| + 1\\s''\neq p,\;
      \mathrm{step}=2}}^{p - 1 - |p - s - s'|} \cchi^{+}(s'') +
  \delta_{p, s, s'}\cchi^{+}(p)\\*
  {}+ \smash[t]{\sum_{\substack{s''= 2p - s - s' + 1\\ 
        \mathrm{step}=2}}^{p - 1}}
  (2\cchi^{+}(s'') + 2\cchi^{-}(p - s'')).
\end{multline}
We first calculate the right-hand side.  Simple manipulations with
$q$-binomial coefficients show that
\begin{multline*}
  \cchi^{+}(s)+\cchi^{-}(p-s) =(-1)^{s+1}
  \smash[b]{\sum_{n=0}^{p-1}\sum_{m=0}^{p-1}}
  (\q-\q^{-1})^{2m}
  \q^{-(m+1)(m+s-1-2n)}\\*
  {}\times\QBIN{s+m-n-1}{m}\qbin{n}{m} E^m F^m K^{s-1 - 2n + m},
\end{multline*}
where
\begin{equation*}
  \QBIN{m}{n}=
  \begin{cases}
    0,& n<0,\\
    \ffrac{[m-n+1]\dots[m]}{[n]!}&\text{otherwise},
  \end{cases}
\end{equation*}
which leads to
\begin{multline*}
  \text{r.-h.\ s.\ of~\eqref{lambda-id}}
  = (-1)^{s+s'}\sum_{m=0}^{p-1 
  } \sum_{\ell=0}^{\min(s, s') - 1}\sum_{n=\ell}^{s + s' - 2 - \ell}
  (\q - \q^{-1})^{2m}
  \q^{-(m + 1)(m + s + s' - 2 - 2n)}\\*
  {}\times\qbin{s + s' - 2 - \ell - n + m}{m} \qbin{n - \ell}{m} E^m
  F^m K^{s + s' - 2 - 2n + m}.
\end{multline*}
Changing the order of summations, using that the $q$-binomial
coefficients vanish in the cases specified in~\eqref{q-bin},
and summing over even and odd $m$ separately, we have
\begin{multline}\label{RHStransf}
  \text{r.-h.\ s.\ of~\eqref{lambda-id}} = \sum_{\substack{m=0\\ 
      \text{even}}}^{p - 1} \sum_{n=0}^{2p - 1}
  \sum_{\ell=0}^{\min(n+\frac{m}{2}, s - 1, s' - 1)}
  (\q - \q^{-1})^{2m}\q^{-(m + 1)(s + s' - 2 - 2n)}\\*
  \shoveright{{}\times(-1)^{s+s'}
    \qbin{s + s' - 2 - \ell - n + \frac{m}{2}}{m}
    \qbin{n +\frac{m}{2} - \ell}{m}
    E^m F^m K^{s + s' - 2 - 2n}+{}}\\
  {}+\smash[b]{
    (-1)^{s+s'}\sum_{\substack{m=1\\ \text{odd}}}^{p - 1}}
  \sum_{n=0}^{2p - 1}
  \sum_{\ell=0}^{\min(n+\frac{m-1}{2}, s - 1, s' - 1)}
  (\q - \q^{-1})^{2m}\q^{-(m + 1)(s + s' - 2n - 1)}\\*
  {}\times \qbin{s + s' - 2 - \ell - n + \frac{m+1}{2}}{m} \qbin{n
    +\frac{m-1}{2} - \ell}{m} E^m F^m K^{s + s' - 2n - 1}.
\end{multline}

Next, in the left-hand side of~\eqref{lambda-id}, we use that
$\cchi^{+}(s)$ are central and readily calculate
\begin{multline*}
  \text{l.-h.\ s.\ of~\eqref{lambda-id}}
  =
  \smash[b]{(-1)^{s+1}\sum_{n=0}^{s-1}
    \sum_{m=0}^{n}}
  (\q-\q^{-1})^{2m} \q^{-(m+1)(m+s-1-2n)}\\*
  \shoveright{{}\times\qbin{s-n+m-1}{m}\qbin{n}{m}
    E^m \cchi^{+}(s') F^m K^{s - 1 - 2n + m}={}}\\
  \shoveleft{{}
    = {(-1)^{s+s'}\sum_{m=0}^{p - 1}
      \sum_{n'=0}^{s' - 1}
      \sum_{n=n'}^{s + n' - 1}     
      \sum_{j=0}^{p-1 
      }}
    (\q - \q^{-1})^{2m} \q^{-m(m + s' - 2n')}
    \q^{-(j + 1)(s + s' - 2 - 2n)}}\\*
  {}\times
  \qbin{s {-} n {+} n' {+} j {-} 1}{j} \qbin{n {-} n'}{j}
  \qbin{s' {-} n' {+} m {-} j {-} 1}{m {-} j}\qbin{n'}{m {-} j}
  E^m F^m K^{s + s' - 2 - 2n + m}.
\end{multline*}
Changing the order of summations, using that the $q$-binomial
coefficients vanish in the cases specified in~\eqref{q-bin},
and summing over even and odd $m$ separately, we have
\begin{multline}\label{LHStransf}
  \text{l.-h.\ s.\ of~\eqref{lambda-id}}
  =\sum_{\substack{m=0\\ \text{even}}}^{p - 1}\!
  \sum_{j=0}^{p - 1}\!
  \sum_{n=0}^{2p - 1}\!
  \sum_{n'=0}^{s' - 1}
  (\q {-} \q^{-1})^{2m} \q^{-m(m + s' - 2n')
    -(j + 1)(s + s' - 2 - 2 n - m)}\\*
  \shoveright{{}\times (-1)^{s+s'}
    \qbin{s {-} n {-} \frac{m}{2}+ n' + j {-} 1}{j}
    \qbin{n {+} \frac{m}{2} {-} n'}{j}
    \qbin{s' {-} n' {+} m {-} j {-} 1}{m {-} j} \qbin{n'}{m {-} j}
    E^m F^m K^{s + s' - 2 - 2n}}\\*
  \shoveleft{{}+(-1)^{s+s'}\sum_{\substack{m=1\\ \text{odd}}}^{p - 1}
    \sum_{j=0}^{p - 1}
    \sum_{n=0}^{2p - 1}
    \sum_{n'=0}^{s' - 1}
    (\q - \q^{-1})^{2m} \q^{-m(m + s' - 2n')}
    \q^{-(j + 1)(s + s' - 2 n - m - 1)}}\\*
  {}\times \qbin{s {-} n {-} \frac{m-1}{2} {+} n' {+} j {-} 1}{j}
  \qbin{n {+} \frac{m-1}{2} {-} n'}{j}
  \qbin{s' {-} n' {+} m {-} j {-} 1}{m {-} j} \qbin{n'}{m {-} j}
  E^m F^m K^{s + s' - 2n - 1}.
\end{multline}

Equating the respective coefficients at the PBW-basis elements
in~\eqref{LHStransf} and~\eqref{RHStransf}, we obtain
\begin{multline*}
  \sum_{j=0}^{p - 1}\sum_{i=0}^{p - 1} \q^{2 m i + j (2 n + 2 - s -
    s')} \qbin{n - i}{j} \qbin{i}{m - j} \qbin{i + j + s - 1 - n}{j}
  \qbin{m - i - j - 1 + s'}{m - j}={}\\
  {}= \q^{m(2 n + 1 - s)} \sum_{\ell=0}^{\min(s - 1, s' - 1)} \qbin{n
    - \ell}{m}\qbin{m + s + s' - 2 - \ell - n}{m},
\end{multline*}
where $1\leq m\leq p-1$, $n\,{\in}\,\oZ_{2p}$, $1\leq s,s'\leq p$.
Because of the vanishing of $q$-binomial coefficients
(see~\eqref{q-bin}), the summations over $j$ and $i$ in the left-hand
side can be extended to $\oZ\times\oZ$, which
gives~\eqref{the-identity} after the shifts $s\to s + 1$, $s'\to s' +
1$.  In the above derivation, $\q$ was the $2p$th primitive root of
unity, but because $p$ does not explicitly enter the resultant
identity and because $q$-binomial coefficients are (Laurent)
polynomials in $q$, we conclude that~\eqref{the-identity} is valid for
all~$q$.\enlargethispage{\baselineskip}

\end{document}